\documentclass[useAMS,usenatbib]{mn2e}

\newcounter{footnew}
\newcommand{\startfoot}{\setcounter{footnew}{0}}
\newcommand{\valfoot}{\stepcounter{footnew}\footnotemark[\value{footnew}]}
\newcommand{\newfoot}{
\valfoot
\renewcommand*{\thefootnote}{\arabic{footnote}}
}
\newcommand{\textfoot}[1]{\newfoot{} {\footnotesize #1 }\\}

\usepackage[english]{babel}

\usepackage{graphicx}
\usepackage{amssymb}
\usepackage{rotating}
\usepackage{tablefootnote}

\usepackage{journals}
\usepackage{amsmath}
\usepackage{natbib}
\usepackage{color}

\newcommand{\figbox}[1]{\makebox[0.235\textwidth]{\includegraphics[width=0.23\textwidth]{#1}}} 
\newcommand{\twofigbox}[2]{\figbox{#1}\hspace{\stretch{0}}\figbox{#2}\\ \vspace{3pt}}
\newcommand{\change}[1]{\textcolor{black}{#1}}

\begin{document}

\voffset= -1.3cm 

\title[NUV companions to millisecond pulsars in 47\,Tuc]{Discovery of near-ultraviolet counterparts to millisecond pulsars in the globular cluster 47\,Tucanae$^1$}
\author[L.E. Rivera-Sandoval et al.]{L. E. Rivera-Sandoval$^{1}$\thanks{E-mail: l.e.riverasandoval@uva.nl}, 
M. van den Berg$^{1,2}$, C. O. Heinke$^{3,4}$, H. N. Cohn$^{5}$, 
\newauthor  P. M. Lugger$^{5}$, P. Freire$^{4}$,  J. Anderson$^{6}$, A. M. Serenelli$^7$, L. G. Althaus$^{8}$, 
\newauthor  A. M. Cool$^{9}$, J. E. Grindlay$^{2}$, P. D. Edmonds$^{2}$, R. Wijnands$^{1}$ and N. Ivanova$^{3}$\\
$^{1}$ Anton Pannekoek Institute for Astronomy, University of Amsterdam, Science Park 904, 1098 XH Amsterdam, The Netherlands\\
$^{2}$ Harvard-Smithsonian Center for Astrophysics, 60 Garden Street, Cambridge, MA 02138, USA\\
$^{3}$ Department of Physics, University of Alberta, CCIS 4-183, Edmonton, AB T6G 2E1, Canada\\
$^{4}$ Max Planck Institute for Radio Astronomy, Auf dem Hugel 69, 53121 Bonn, Germany\\
$^{5}$ Department of Astronomy, Indiana University, 727 E. Third St, Bloomington, IN 47405, USA\\
$^{6}$ Space Telescope Science Institute, 3700 San Martin Drive, Baltimore, MD 21218, USA\\
$^{7}$ Instituto de Ciencias del Espacio (CSIC-IEEC), Facultad de Ciencias, Campus UAB, 08193, Bellaterra, Spain\\
$^{8}$ Instituto de Astrof\'isica La Plata, Facultad de Ciencias Astron\'omicas y Geof\'isicas (CONICET-UNLP)\\
\ \ \ Paseo del Bosque s/n, 1900 La Plata, Argentina\\
$^{9}$ Department of Physics and Astronomy, San Francisco State University, 1600 Holloway Avenue, San Francisco, CA 94132, USA}


\pagerange{\pageref{firstpage}--\pageref{lastpage}} \pubyear{2015}

\maketitle

\label{firstpage}

\begin{abstract}

\noindent
We report the discovery of the likely white dwarf companions to radio millisecond pulsars 47\,Tuc Q and 47\,Tuc S
in the globular cluster 47\,Tucanae. These blue stars were found in near-ultraviolet
  images from the {\em Hubble Space Telescope} for which we derived
  accurate absolute astrometry, and are located at positions
  consistent with the radio coordinates to within $0\farcs016$ ($0.2\,\sigma$). 
We present near-ultraviolet and optical colors for the previously identified companion to millisecond pulsar 47\,Tuc U, and
we unambiguously confirm the tentative prior identifications of the optical counterparts to 47\,Tuc  T and 47\,Tuc  Y.
For the latter we present its radio timing solution for the first time.
We find that all five near-ultraviolet counterparts have
    $U_{300}-B_{390}$ colors that are consistent with He white dwarf
    cooling models for masses $\sim0.16-0.3$ $M_{\odot}$ and cooling
    ages between $\sim0.1-6$ Gyr. The H$\alpha-R_{625}$ colors of
    47\,Tuc U and 47\,Tuc T indicate the presence of a strong H$\alpha$
    absorption line, as expected for white dwarfs with a H envelope.

\end{abstract}

\begin{keywords}
Binaries: general - globular clusters: individual (47\,Tucanae) - pulsars: general - pulsars: individual (PSR J0024--7204Q, PSR J0024--7204S, PSR J0024--7204T, 
PSR J0024--7203U, PSR J0024--7204Y)
\end{keywords}

\section{Introduction}

\begin{table*}
\label{table:radio}
 \centering
 \startfoot   
  \caption{Parameters of the millisecond pulsars in 47\,Tuc whose companions are discussed in this paper, for a reference epoch MJD 51600. 
Positions, spin periods ($P_s$), orbital periods ($P_b$), limits on companion masses ($m_c$; for an assumed neutron star mass of 1.35 $M_{\odot}$), and characteristic ages ($\tau_c$) 
are from Freire et al. (in preparation). X-ray luminosities in the $0.1-10$ keV band ($L_X$) were taken from \citet{2006-bog}. \change{Units of right ascension are hours, minutes, and seconds, and units of declination are degrees, arcminutes, and arcseconds.}
Numbers in parentheses are $1\sigma$ uncertainties in the last significant digit, errors in $P_s$ are in thirteenth decimal, while errors in $P_b$  are in the ninth decimal or smaller.
Values of $\tau_c$ are lower limits based on the 2\,$\sigma$ upper limit for the spin-period derivatives.}
  \begin{tabular}{c l l c c c c c c}
  \hline
   MSP & R.A. (J2000) & Decl. (J2000) & $P_s$ & $P_b$ &  $m_{c}$ & $\tau_c$ & $L_{X}$ \\
&&&(ms)&(days)&($M_\odot$) &(Gyr)&$(10^{30}$ erg s$^{-1})$\\
 \hline
Q& 00:24:16.4903(2) & --72:04:25.1653(8) & 4.03& 1.189& $>0.18$&$>1.43$&3.9\\
S&00:24:03.9794(1) &--72:04:42.3535(5)& 2.83& 1.201 & $>0.09$&$>0.91$&8.9\\
T&00:24:08.5487(7) &  --72:04:38.928(3) & 7.58 & 1.126 & $>0.17$&$>0.32$&2.9\\
U& 00:24:09.83626(7) & --72:03:59.6889(3) & 4.34&0.429 &  $>0.12$&$>1.91$&5.1\\
Y&00:24:01.4016(3) & --72:04:41.837(1) & 2.19 &0.521 & $>0.14$ &$>2.2\phantom{0}$&3.9  \\
\hline
\end{tabular}
\vspace{-3pt}  
{\flushleft
\startfoot  
 } 
\end{table*}

\stepcounter{footnote}
Millisecond pulsars (MSPs) are rapidly spinning neutron stars with\footnotetext{\change{Based on proprietary and archival observations with the NASA/ESA Hubble Space Telescope, obtained at the Space Telescope Science Institute, which is operated by AURA, Inc., under NASA contract NAS 5-26555.}}
spin periods of 20 ms or less, magnetic field strengths $B \approx
10^{8-9}$ G, and characteristic spin-down ages $\tau_c \approx
10^{9-10}$ yr \citep[see e.g.\, the review by][]{2008-Ransom}. Their properties can be explained by the
recycling formation scenario \citep{1982-alpar}, in which an old
neutron star is spun up by the accretion of mass from a
companion star in a low-mass X-ray binary (LMXB) phase.

Up to now, more than 200 MSPs have been discovered in the Galaxy, and
a large fraction of these reside in globular clusters (GCs). The number of
MSPs per unit mass is higher in GCs than in the Galactic
disk by two to three orders of magnitude \citep{2013-Freire}. In the context of the recycling scenario, this overabundance
is no surprise. The LMXB progenitors of MSPs have long been known to
occur in GCs with a much higher frequency than in the
rest of the Galaxy \citep{1975-Clark} as a result of the high stellar
encounter rates that favor their formation in dense clusters.

Most MSPs are found in binary systems. Studies of the orbital
parameters and the companion properties can provide valuable
information on the advanced evolution of LMXBs, the pulsar spin-up
process \change{\citep[e.g.][]{1976-Smarr}}, and (for MSPs in GCs) the effects of dynamical
encounters on the characteristics of the MSP population 
\citep[see e.g.][]{2014-Ver}. Various classes of binary MSPs can be
distinguished; each of these are found in the Galactic field and in
GCs. The systems that are expected to result from the
standard recycling scenario have low-mass (a few tenths of 
$M_{\odot}$) 
white dwarf (WD) companions that descended from the original
mass donors in the LMXBs. Most MSPs are found in this
configuration. The very-low mass binary MSPs (or black widows) likely
represent the very latest stages of evolution, in which the companion
(with only a few hundredths of $M_{\odot}$ in mass remaining) is being
ablated by the pulsar. Finally, the so-called redback MSPs are
characterized by radio eclipses and non-degenerate companions of $0.2-0.4$ $M_{\odot}$
(see \citealt{2013-Roberts} for a review). These may be systems in transition
between an LMXB phase and a detached phase where the MSP can be
detected as a radio pulsar, or (in globular clusters) the result of
exchange encounters.

The identification of MSPs at other wavelengths enables a range of
studies into the physical processes that govern the evolution of the
MSP binaries. In GCs, the spatial resolution of the {\em Chandra X-ray
  Observatory} and the {\em Hubble Space Telecope} ({\em HST}) is a
necessity to overcome the crowding. 
Indeed, the first unambiguous
X-ray identification of MSPs in globular clusters was made with the
first {\em Chandra} observation of 47\,Tuc, in which fifteen of the 23 MSPs
then known were first detected and precisely located in X-rays
\citep{2001-grindlay}. 
So far, {\em HST} counterparts to
globular-cluster MSPs have been found for only twelve systems, two of
which are tentative identifications \citep[47\,Tuc T and
  Y;][]{2003-edmonds1}.  The other ten systems in GCs are 47\,Tuc U
and W (\citeauthor{2001-edmondsw11}
\citeyear{2001-edmondsw11,2002-Edmonds}), NGC 6397 A
(\citeauthor{2001-ferrarob} \citeyear{2001-ferrarob}), NGC 6752 A
(\citeauthor{2003-Bassa} \citeyear{2003-Bassa},
\citeauthor{2003-ferraro} \citeyear{2003-ferraro}), M4 A
(\citeauthor{2003-Sirgu} \citeyear{2003-Sirgu}), NGC 6266 B
(\citeauthor{2008-cocozza} \citeyear{2008-cocozza}), M28 H and I
(\citeauthor{2013-Pallanca} \citeyear{2010-pallanca},
\citeyear{2013-Pallanca}), M5 C (\citeauthor{2014-Pallanca}
\citeyear{2014-Pallanca}) and M71 A (\citeauthor{2015-cadelano}
\citeyear{2015-cadelano}).  Clearly, more identifications are needed
to get a better understanding of the possible differences between MSP
properties in GCs and in the field.

Right after Terzan\,5, 47\,Tucanae (47\,Tuc) is the GC
where most radio MSPs have been identified so far: a total of 23 were
discovered \citep{1990-Manchester,1991-Manchester,1995-robinson,2000-camilo}, 
fifteen of which are in binary
systems\footnote{http://www.naic.edu/$\sim$pfreire/GCpsr.html}. 
To date, accurate radio-timing positions have been
derived for thirteen of the 47\,Tuc binary MSPs \citep[][and in preparation]{2003-freire}, 
which allows a search for their counterparts at
ultraviolet and optical wavelengths. So far, secure identifications
have been found for only two of them. The companion of PSR J0024--7204W, or 
47\,Tuc W in short, is a
non-degenerate star that may be (close to) filling its Roche lobe
\citep{2002-Edmonds,2005-bogdanov}; it was among the first
MSP binaries to be classified as a "redback" system. The companion of
47\,Tuc U, on the other hand, is typical for a system that is the
outcome of the canonical recycling scenario. Edmonds et al.(2001; E01 hereafter)
 found that it is likely a He WD\footnote{
He WDs are low-mass ($\lesssim 0.45$ $M_{\odot}$) white dwarfs with He-rich cores. 
They are thought to have formed as a result of a mass-transfer episode in a close binary 
before He ignition, removing much of the He WD-progenitor's envelope and exposing its He-rich core.} 
with a mass of $\sim$0.17
$M_{\odot}$.  \citeauthor{2003-edmonds1} (2003a; from now on E03a) 
\defcitealias{2003-edmonds1}{E03a} suggested identifications
for two more systems, 47\,Tuc Y\footnote{The optical counterpart to 47\,Tuc Y was originally
    reported as the counterpart to the {\em Chandra} source W\,82 by
    \citetalias{2003-edmonds1}. This source was classified as a candidate cataclysmic
    variable, as the radio position for 47\,Tuc Y was unknown at the
    time. After the position became available, Bogdanov et al. 2006
    reported its association with W\,82, implying the detection of
    the MSP companion by E03a.}
 and T, but from their data they could only derive upper limits on
the $U-V$ colors of the proposed blue counterparts.
\nocite{2001-edmondsw11}

We have carried out a search for counterparts to the 47\,Tuc MSPs
using new near-ultraviolet (NUV) imaging data obtained with the {\em
HST} Wide Field Camera 3 (WFC3), and optical data (including deep H$\alpha$ imaging)
from the Advanced Camera for Surveys (ACS). The NUV data are particularly useful
for looking for WD companions, whose blue spectral energy
distribution yields a high contrast against the typically red cluster
stars. Here we present the results of our search, which includes the
discovery of the counterparts to 47\,Tuc Q and S, and the
corroboration of the suggested identifications of 47\,Tuc T and Y. We
find that, like in 47\,Tuc U, the MSP companion in all four systems is
likely a low-mass ($0.16-0.3$ $M_{\odot}$) He WD.

This paper is structured as follows. In Section 2 we describe the {\em HST} observations, and the X-ray data used in this work. 
The {\em HST} data analysis 
is described in Section 3. In Section 4 we present 
the search for counterparts to the 47\,Tuc MSPs in the {\em HST}
data, and we show our results
for the new identifications and the known WD companion of 47\,Tuc U. Finally
we discuss our results in Section 5.

\section[]{Observations}

The NUV data used in this work\change{\footnote{\change{We have not analyzed the existing STIS far-ultraviolet data of the center of 47\,Tuc, because the five objects studied in this paper are not included in the small field of view of that data set.}}} belong to {\em HST} program GO $12950$ taken with the UVIS channel of the WFC3. The data set consist of eight orbits divided in two consecutive visits on 2013 August 13 starting at $02$:$10$ UTC and finishing at $11$:$01$ UTC of the same day. 
Each visit started with one orbit of four F390W exposures of about 580 s each. This was followed by three orbits of twelve F300X exposures in total, with individual exposure times of about 609 s, resulting in a total exposure time of $3881$ s in the F390W filter and $14256$ s in the F300X filter. The native scale of the WFC3/UVIS images is $0\farcs04$ pixel$^{-1}$. 
The UVIS detector is composed of 2 chips, having a chip gap between them equivalent to $1\farcs2$ on the sky. In order to improve our spatial resolution, we used a $4-$point dithering pattern that includes fractional-pixel offsets. In both visits an offset between images was imposed to ensure coverage of the chip gap.
The images were centered near the center of 47\,Tuc as reported by \citet{2011-goldsbury}. Given that the WFC3/UVIS has a field of view \change{(FOV)} of $162''\times162''$, our images only cover $\sim 23\%$ of the area inside the half-mass radius of the cluster (r$_h=3\farcm17$, \citeauthor{1996-harris} \citeyear{1996-harris}, 2010 edition\footnote{Catalog of parameters for Milky Way GCs, 
http://www.physics.mcmaster.ca/Fac$\_$Harris/mwgc.dat})

The optical observations reported here were taken with the Wide Field Channel (WFC) on the ACS under program
GO\,9281. We obtained images in three filters: F435W,
F658N, and F625W, which encompasses H$\alpha$
and the surrounding continuum. The observations were spread over three
visits on 2002 September 30 (01:25--06:31 UTC), October 2/3
(20:37--00:51 UTC), and October 11 (02:04--10:40 UTC). The visits were
scheduled to be simultaneous with the deep 47\,Tuc {\em Chandra}
observations that are presented in \citeauthor{2005-heinkea} (\citeyear{2005-heinkea}; hereafter H05).
The total
exposure time amounted to 955 s in F435W, 1320 s in F625W, and 7440 s
in F685N. Observations through all three filters were taken in each
visit, resulting in near-simultaneous colors. Since X-ray emitting
binaries frequently display variability in X-ray and optical
wavelengths, this provides an optimal estimate of the system's actual
colors, which is unaffected by long-term variability in one or both
wavelength bands. The FOV of the WFC is
202\arcsec$\times$202\arcsec, imaged with two CCD detectors separated
by a 2\farcs45-wide gap. We used a dither pattern with fractional-pixel offsets
to be able to improve the native WFC pixel scale of 0$\farcs$05 pixel$^{-1}$
during the image stacking.  A complete description of the
GO 9281 data and results is forthcoming (van den Berg et al.~in
preparation). Of the twenty MSPs in 47\,Tuc with radio-timing positions, nineteen are covered by the UVIS and WFC images;
only 47\,Tuc X, a binary MSP, was missed. 

In this work we make use of the catalog of X-ray sources in 47\,Tuc
    from H05 (see that paper for details about the data
    reduction), which was obtained from deep observations (281 ks in
    total) with {\em Chandra}. All nineteen MSPs covered by our NUV/optical data are matched to an X-ray source in this catalog.

\section[]{Data reduction and analysis}

The data reduction of our UVIS data starts with the flatfielded
    images produced by the standard UVIS pipeline (CALWF3 version 3.1.2). To correct for the charge transfer efficiency
    (CTE) degradation of the UVIS detectors, we used the CTE
    correction software\footnote{http://www.stsci.edu/hst/wfc3/tools/cte\_tools} provided by the Space Telescope Science Institute
    (STScI). The resulting images still suffer from geometric distortion, which
we corrected for with the DrizzlePac\footnote{http://drizzlepac.stsci.edu} software (Gonzaga et al. \citeyear{2012-Gon}). Astrometric alignment of
  the individual images was performed with the TWEAKREG task in
  Drizzlepac. Subsequently, the ASTRODRIZZLE task was used to combine
  the aligned images into one stacked master frame in each filter that
  is distortion free, and cleaned of cosmic rays and detector
  artifacts like hot pixels or bad columns. The resulting master
  images in F390W and F300X are twice-oversampled to a pixel scale of
  0\farcs02 pixel$^{-1}$.

The reduction steps for the WFC images are similar to those for
  the UVIS data, except that the images produced by the ACS calibration pipeline
    (CALACS version 8.2.0) are already corrected for CTE losses.
The stacked master images in F435W, F658N,
  and F625W have a final pixel scale of 0\farcs025 pixel$^{-1}$.

\begin{table*}
\centering
 \caption{Astrometric and photometric results for the MSP companions
    discussed in this paper. Errors in the absolute astrometry of the
    celestial positions are $0\farcs074$ (1$\,\sigma$). \change{Units of right ascension are hours, minutes, and seconds, and units of declination are degrees, arcminutes, and arcseconds.} Columns 4 to 8
    give the calibrated photometry on the Vega-mag system and the errors from DAOPHOT. 
    \change{The projected distance of the companions from the cluster center ($\delta_c$) was calculated using the coordinates of the center of 47\,Tuc as reported by \citet{2011-goldsbury}.}
    The offsets between the X-ray and NUV positions (last column) were
    determined using the positions from H05. The value of
    $\sigma_X$ is the combined error in the NUV and X-ray positions.}
\scalebox{0.85}{
  \begin{tabular}{c c c c c c c c c c c c}
  \hline
MSP & R. A. & Decl. & $U_{300}$ & $B_{390}$ &$B_{435}$ & $R_{625}$ &H$\alpha$ & \change{$\delta_c$} &Radio& X-ray$^1$&X-ray \\
companion&(J2000)&(J2000)&&&&&&\change{(arcsec)}&offset& source& offset \\
&&&&&&&&&(arcsec/$\sigma$)&&(arcsec$/\sigma_X$)\\
 \hline
 $\text{Q}_{\rm UV}$ & 00:24:16.493 & --72:04:25.15 & 22.95(3)  &  23.41(4) & 23.84(6)  &  23.6(1) &   22.93(8)&56.89 & 0.012/$0.2$ &W\,104 &$0.05/0.6$\\
 $\text{S}_{\rm UV}$& 00:24:03.976& --72:04:42.34   & 23.16(3)  &  23.77(5) & 23.47(5)  &   --     &    --     &13.09 & 0.016/$0.2$ &W\,77 &$0.34^2/5.7$\\
 $\text{T}_{\rm UV}$ & 00:24:08.551 &--72:04:38.92  & 23.04(2)  &  23.72(3) & 23.9(1)   &  23.5(2) &  23.4(1)  &19.02 &  0.011/$0.1$&W\,105 &$0.25/2.5$\\
$\text{U}_{\rm UV}$& 00:24:09.836  & --72:03:59.69  & 20.545(9) &  20.98(1) & 20.860(5) &  20.82(1)&  20.84(1) &56.33 &  0.002/$0.03$&W\,11 &$0.07/0.9$\\
$\text{Y}_{\rm UV}$ & 00:24:01.402 & --72:04:41.84  & 22.09(2)  &  22.51(5) & 22.36(5)  &  --      &      --   &22.66&0.002/$0.03$&W\,82 &$0.21/1.9$\\
\hline
\end{tabular}
}
\label{tab:offset}
{\flushleft
\startfoot 
\textfoot{Using the nomenclature in H05.}
\textfoot{The X-ray position used for 47\,Tuc S corresponds to the position of a single detected X-ray source that is the combination of X-ray emission
from 47\,Tuc F and 47\,Tuc S. This explains the apparent large X-ray astrometric offset.}
}
\end{table*}

\subsection{Astrometry}
\label{astrometry}

The radio coordinates of the 47\,Tuc MSPs were derived using
 the Solar System ephemeris DE/LE 405. This is 
specified in the International Celestial Reference Frame (ICRF), 
which is aligned to the 
 International Celestial Reference System (ICRS). 
To facilitate the identification of NUV/optical
  counterparts we also aligned our UVIS and WFC images to the ICRS
  using stars in the UCAC2 catalog  \citep{2004-Zaca}. Since
  isolated and unsaturated UCAC2 stars are scarce in the {\em HST}
  images of the cluster core, we derived secondary astrometric
  standards from a ground-based image of 47\,Tuc as an intermediate
  step. We obtained from the ESO archive a 30-s $V$ image taken with
  the 2.2m/Wide Field Imager (WFI) at La Silla, Chile on 2002 October
  29. An astrometric solution of the part of the image that contains
  the cluster center was derived using 225 UCAC2 stars; fitting for
  zero-point offset, rotation, pixel scale, and distortions gives rms
  residuals of about 0\farcs035 in right ascension and declination. We
  then selected 126 secondary standards from the WFI image to
  calibrate a 10 s F435W exposure from our GO\,9281 program, which
  resulted in rms residuals of about 0\farcs027 in each
  direction. Finally, this solution was transferred to our stacked
 WFC images with negligible errors using $>$10\,000 stars,
  and to our stacked UVIS images with rms residuals of about
  0\farcs027 in each direction using $>$11\,700 stars. The final
  1\,sigma error in the absolute astrometry of our images is the
  quadratic sum of all the aforementioned errors, plus a term that
  represents the error with which the UCAC2 is tied to the ICRS. The
  latter is dominated by the estimated systematic error in UCAC2
  positions of 10 mas \citep{2004-Zaca}. For the WFC and UVIS
  images this results in an error of 0\farcs064 and 0\farcs074,
  respectively. Errors in the radio positions are negligible. 

H05 aligned the astrometry of their X-ray source catalog to the radio
positions of seventeen MSPs detected by {\em Chandra}. Therefore, we
can indirectly check the alignment between our NUV and radio positions
by comparing the NUV and {\em Chandra} astrometry. We have done this
by comparing the NUV and X-ray positions of seventeen sources with
published finding charts in E03a and \citeauthor{2003-Edmonds2} (2003b). 
We find that the NUV/X-ray offset is very small, 
about --0\farcs023$\pm$0\farcs008 in right ascension and 
--0\farcs003$\pm$0\farcs009 in declination in both filters, so about one oversampled pixel.

\subsection{Photometry}
\label{Photometry}

Given that the images of the core of 47\,Tuc are very crowded, 
we carried out point spread function (PSF) photometry for the stacked UVIS and WFC frames
by using standard IRAF and DAOPHOT tasks. \change{We used isolated stars to build a variable PSF
for each filter. The obtained PSF was fitted to all detected stars in each stacked frame.} 
\change{Photometric calibration of the extracted magnitudes was performed onto the
 Vega-mag system using the zero points given at STScI's web pages\footnote{http://www.stsci.edu/hst/wfc3/phot\_zp\_lbn http://www.stsci.edu/hst/acs/analysis/zeropoints}; 
 for UVIS we used the zero points provided for $0\farcs4$ apertures.} 
In the remainder we denote calibrated magnitudes in the F300X,
  F390W, F435W, F625W, and F658N filters by $U_{300}$, $B_{390}$,
  $B_{435}$, $R_{625}$, and H$\alpha$, respectively.
The absolute magnitudes were calculated for an adopted distance modulus $(m-M)_0=13.36\pm0.02\pm0.06$,
corresponding to the 47\,Tuc distance of $4.69\pm0.04\pm0.13$ kpc \citep{2012-woodley}. The first error term corresponds to the random error, and the second
to the systematic error of the distance determination.

The UVIS data were also independently analyzed using software based on the program developed for the \emph{ACS Globular Cluster Treasury Project}, 
described in \citet{Anderson08}, which gave consistent results with the DAOPHOT analysis. The reductions were done using the latest version 
of this software known as KS2. \change{Star finding required that stars be detected in both the F300X and F390W frames. Photometry was performed by PSF fitting 
using library PSFs with perturbations to better match the specific PSF in each exposure.} The final KS2 photometry comes from averaging the fluxes from the 
individual images, with sigma clipping applied to remove outlying values due to cosmic rays, defective pixels, etc. A total of 148717 stars were detected in the 
region covered by the UVIS imaging. \change{Color-magnitude diagrams were constructed to choose the best KS2 options for photometry of the faint stars (e.g. the fitting radius), 
from the tightness of the fiducial sequences, and to identify the counterparts for the MSPs.} Drizzle-combined mosaic images of this region were also produced using the STScI PyRaF routine 
ASTRODRIZZLE from the Drizzlepac package. The drizzle-combined images were oversampled by a factor of two, in order to increase the effective resolution. 
Calibration of the KS2 photometry to the Vega-mag system was performed by doing aperture photometry on moderately bright, isolated stars in the image mosaics 
within a $0\farcs1$ radius aperture, finding the aperture correction to an infinite radius aperture from 
ISR WFC3 2009-38\footnote{\change{http://www.stsci.edu/hst/wfc3/documents/ISRs/WFC3-2009-38.pdf}}, 
calculating the median offset between the KS2 photometry and the aperture photometry for these stars, 
and applying the zeropoints from the {\em HST} WFC3 zero point website\footnotemark[9].

To take into account the effect of extinction on the UVIS photometry, we converted the reddening towards 47\,Tuc of 
$E(B-V)=0.04\pm0.02$ \citep{2007-salaris} to the extinction in the F300X and F390W filters using the UVIS 
Exposure Time Calculator of {\em HST}. We obtained extinction values of A(F300X)=0.26$\pm 0.13$ and A(F390W)=0.18$\pm0.09$, respectively. 
To compute the extinction in the WFC filters, we used the conversions in \citet{siriea05}. 
This yields A(F435W)=0.16$\pm$0.08, A(F658N)=0.10$\pm$0.05, and A(F625W)=0.11$\pm$0.05.

\begin{figure}
\centering
\twofigbox{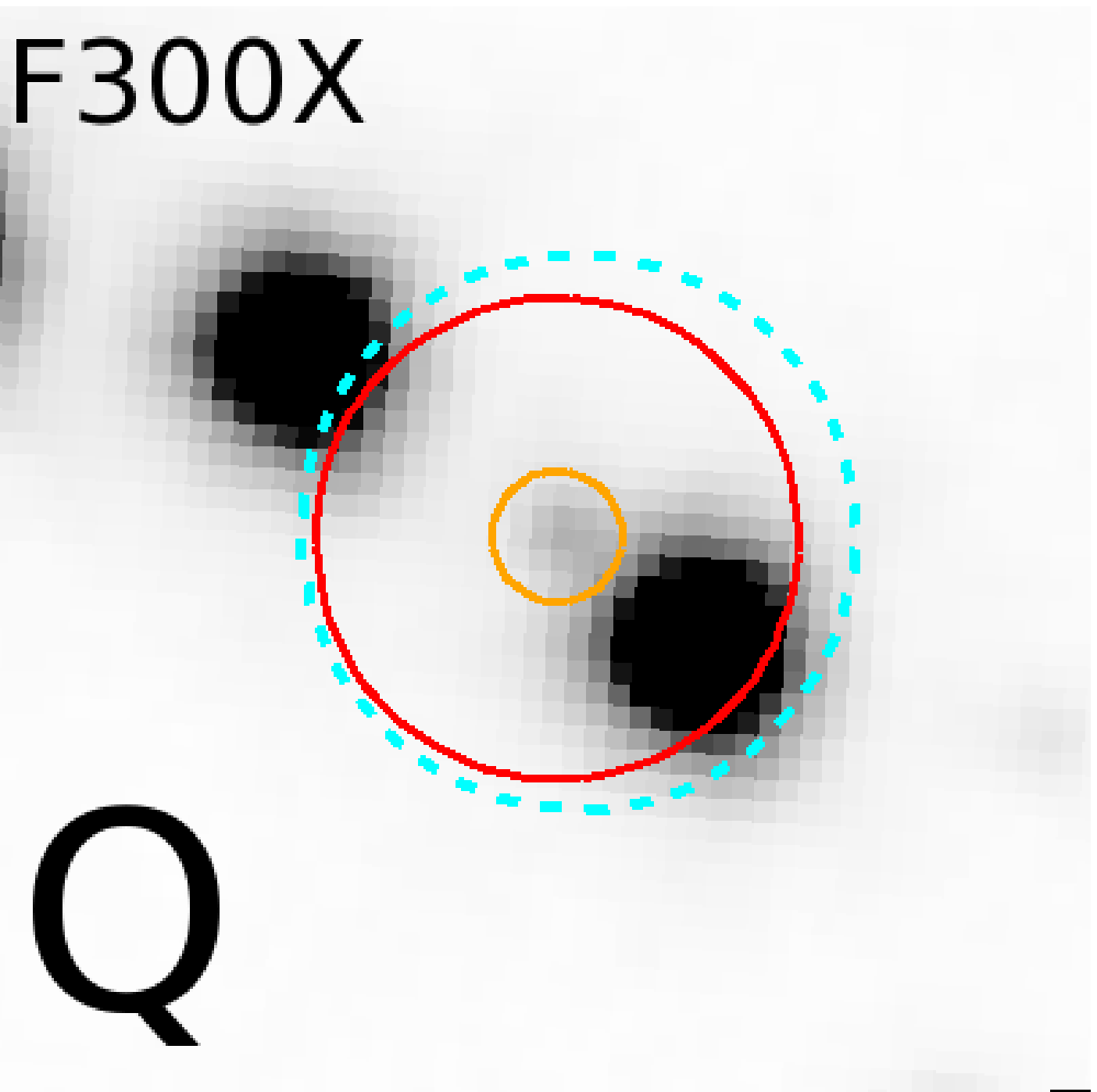}{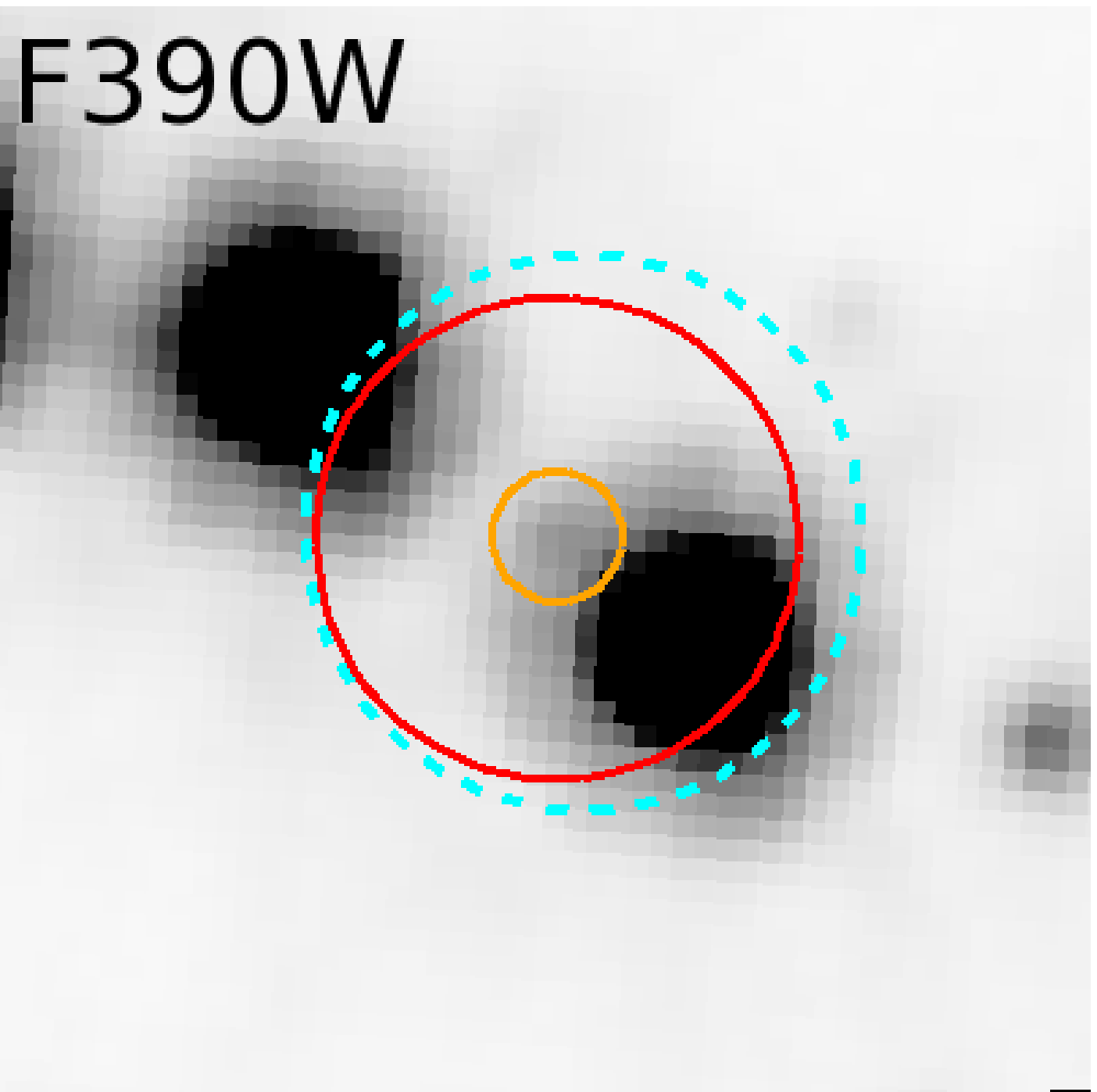}
\twofigbox{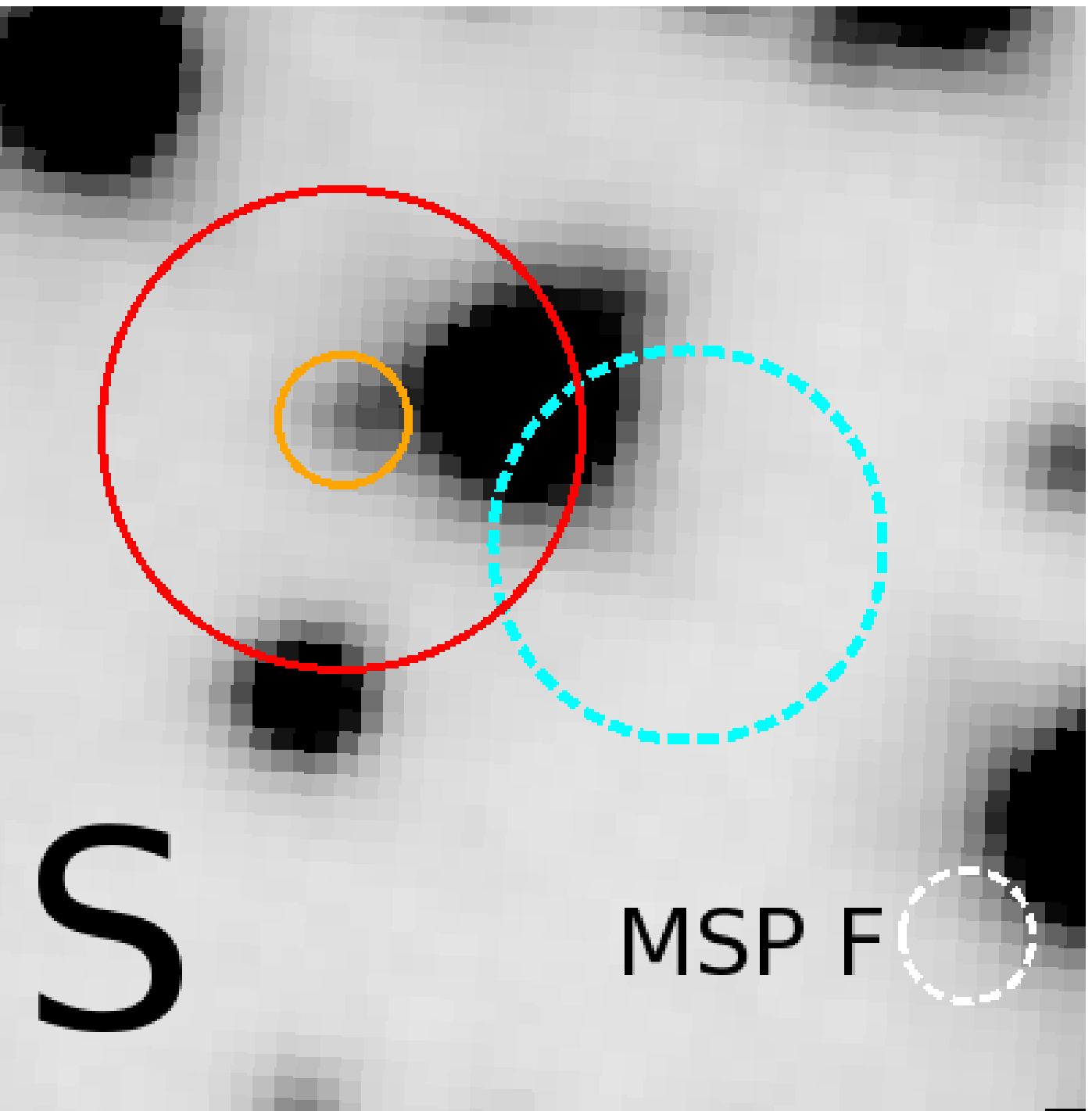}{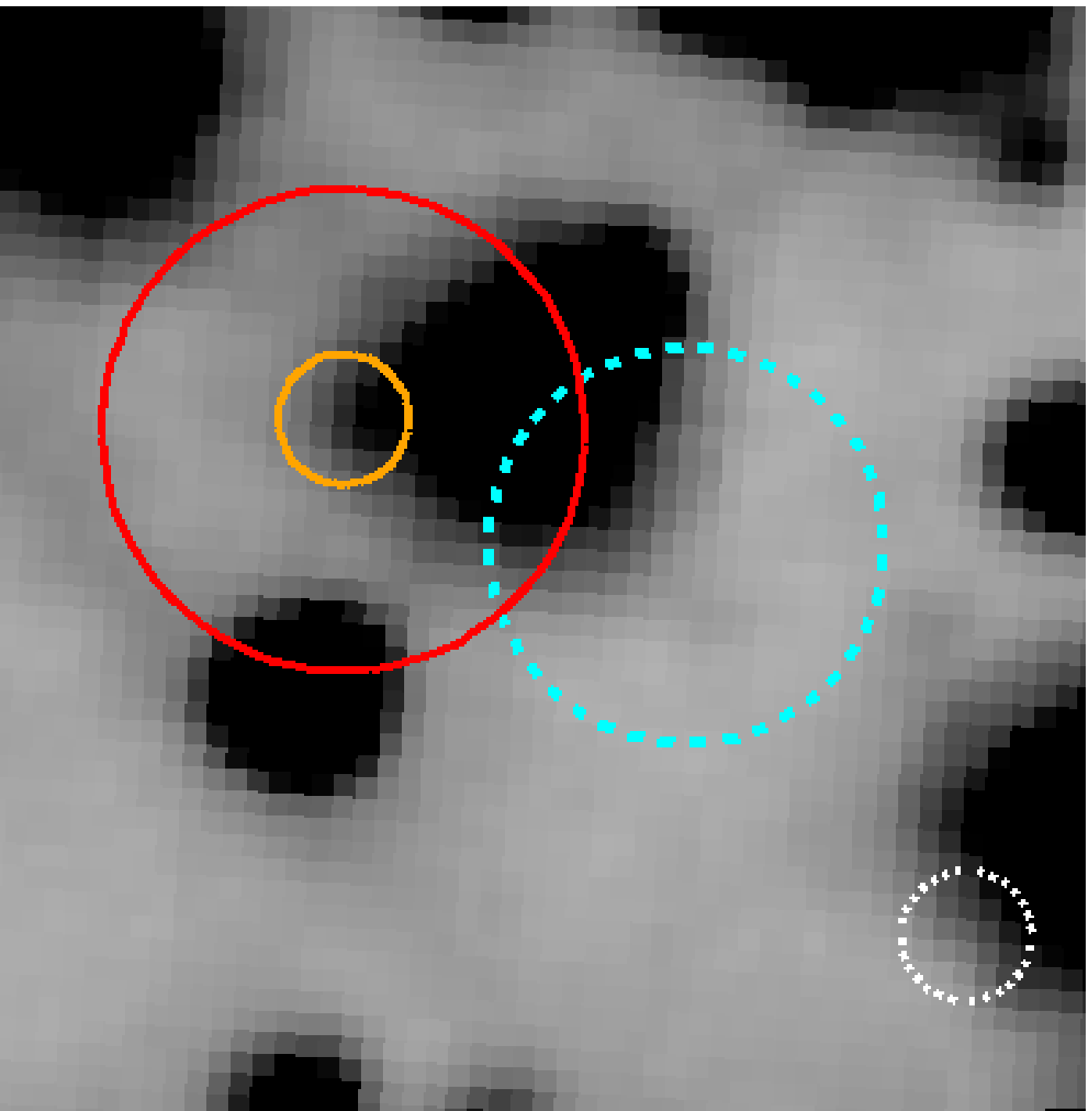}
\twofigbox{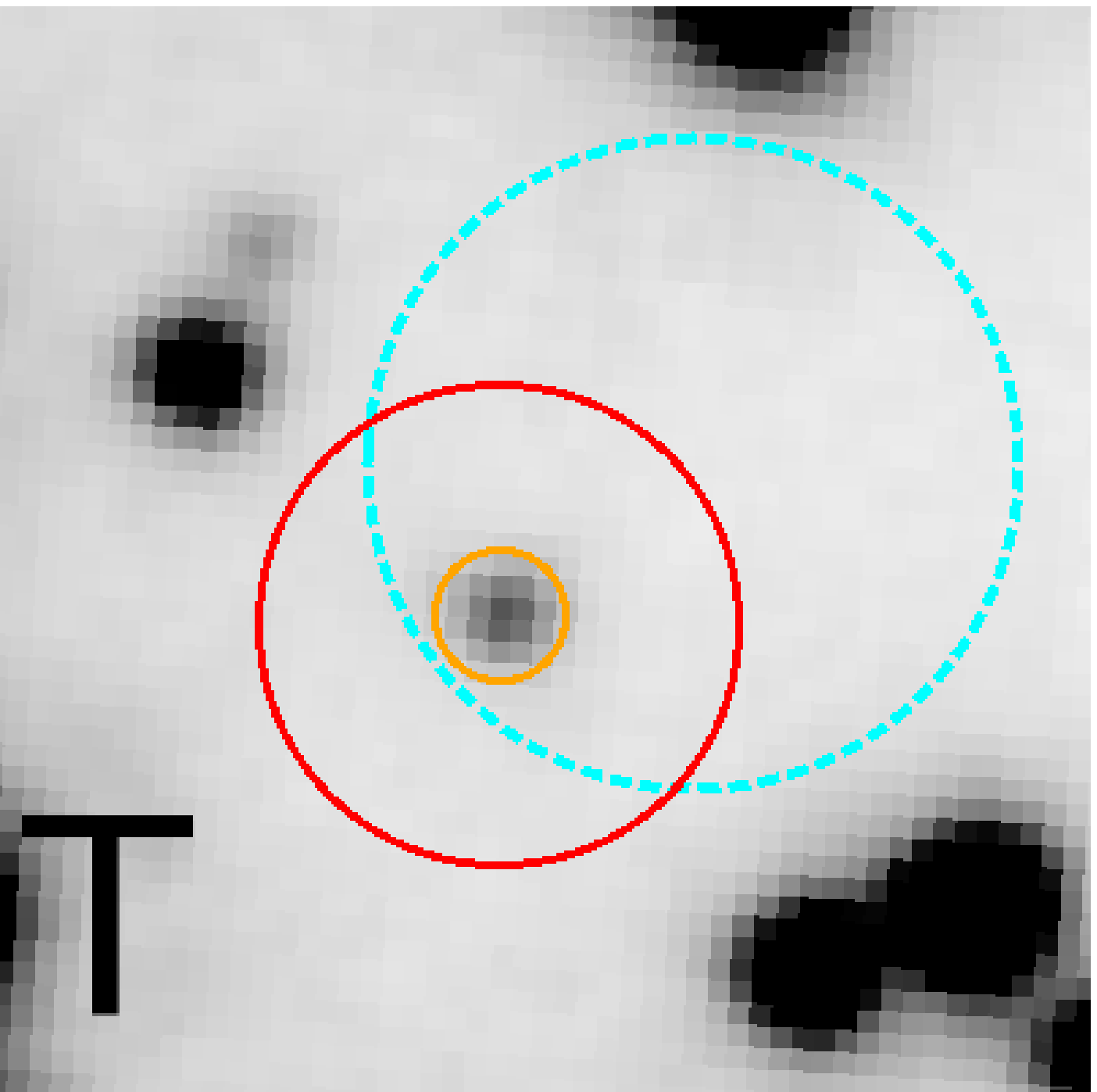}{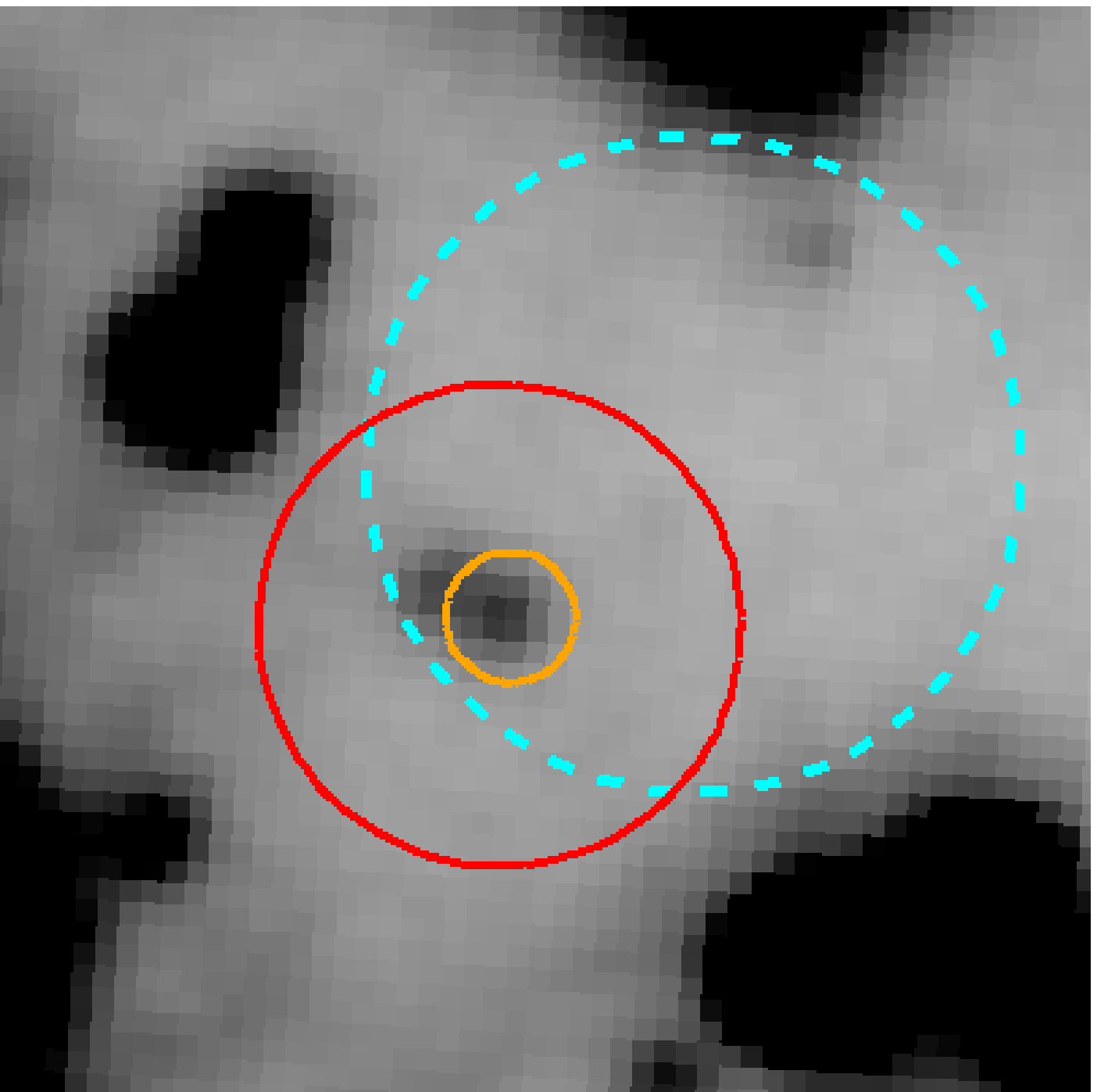}
\twofigbox{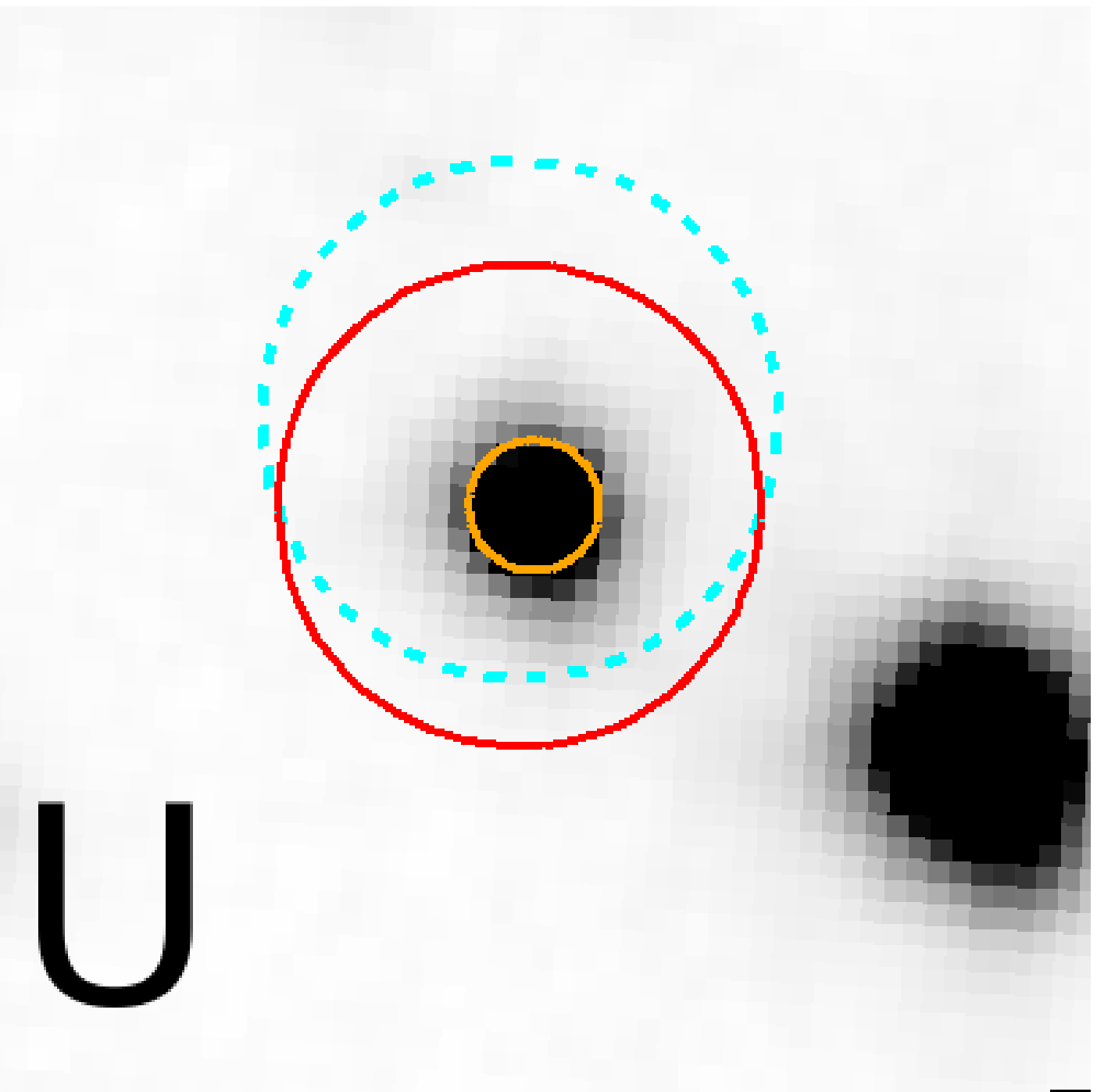}{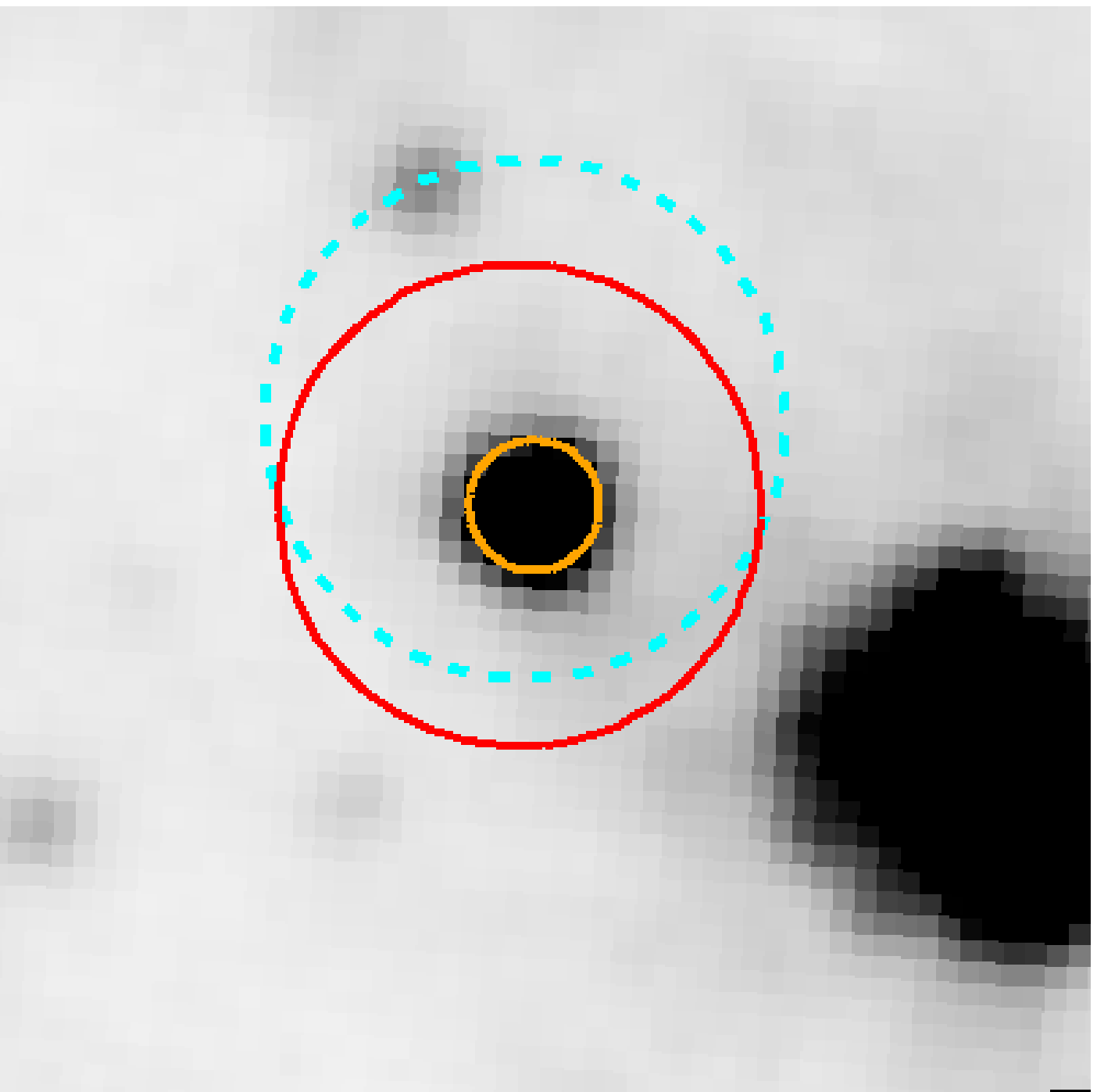}
\twofigbox{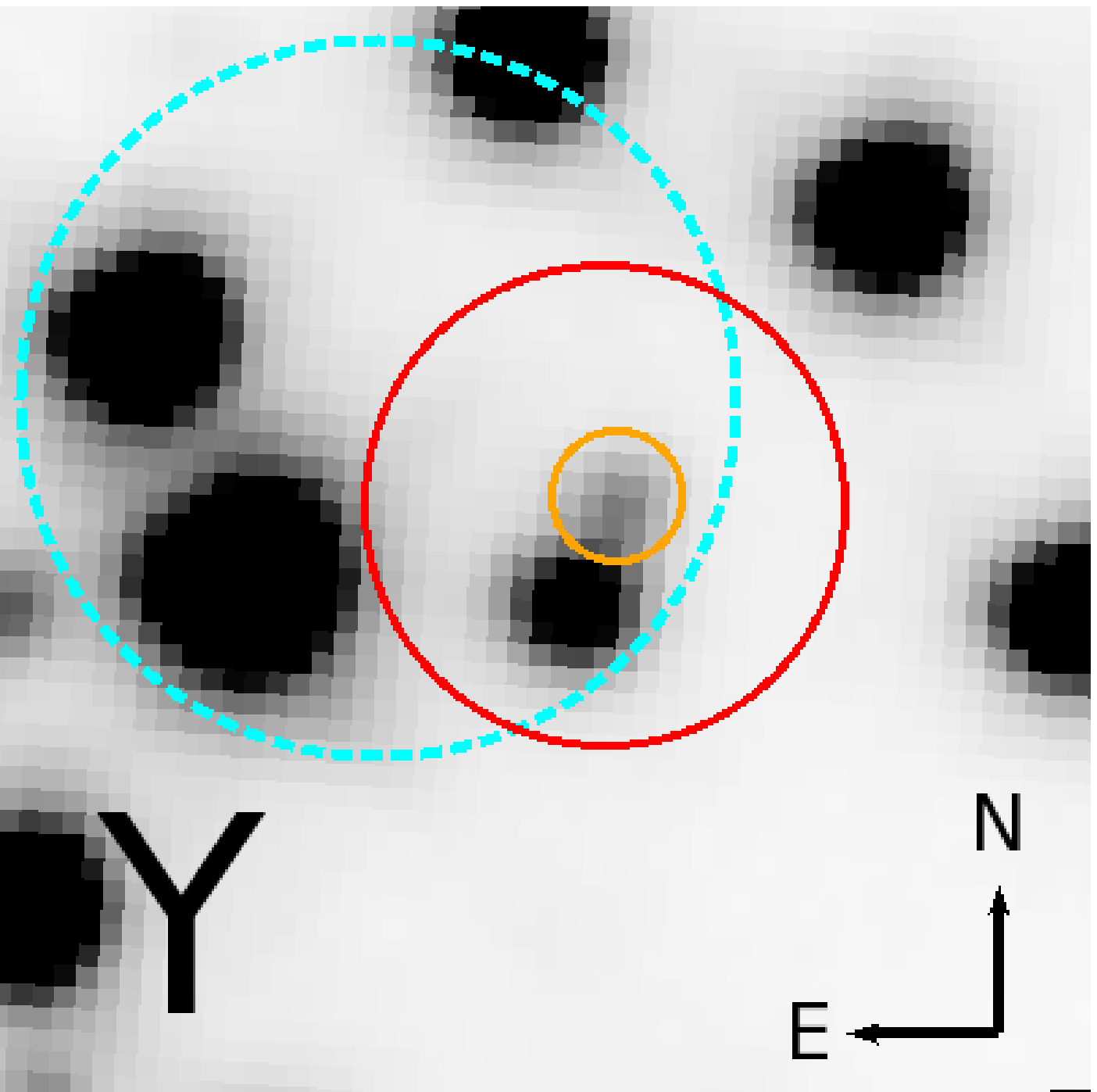}{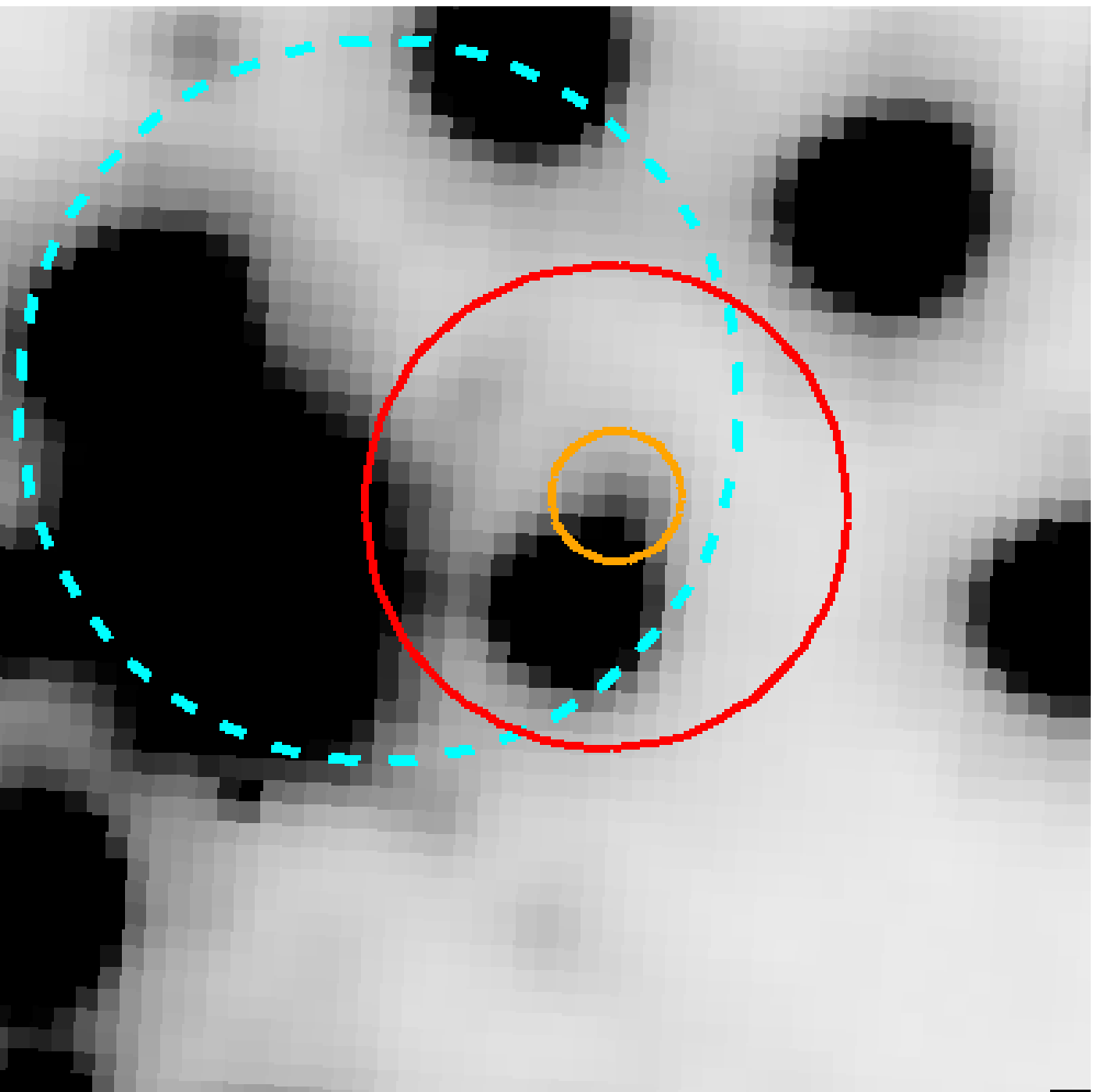}
\caption{Finding charts of the companions to MSPs 47\,Tuc Q, S, T, U and  Y. Images in the filter F300X are shown on the left side while images in F390W are shown on the right. The solid red lines show the 3\,$\sigma$ match circles centered on radio positions of the pulsars, which take into account the astrometric UVIS errors. The dashed cyan circles are the 3\,$\sigma$ match circles of {\em Chandra} sources. The companion candidates are indicated by solid orange circles. Each image is $1\farcs0\times1\farcs0$ in size.}
\label{fig:findings}
\end{figure}

\section[]{Results}
\label{results}

The search for MSP counterparts was done in the astrometrically
  calibrated NUV images. These frames are less crowded than the
  optical images, and contain fewer bright and saturated stars that
  could affect the detection of faint counterparts. We looked for
  counterparts within 3\,$\sigma$ (0\farcs22) of the radio
  positions. 

  We found blue stars inside the match circles of the binary MSPs
  47\,Tuc Q, S, T, U, and Y. These objects are excellent astrometric
  matches with the radio positions, with offsets that are smaller than
  0\farcs016 (0.2\,$\sigma$). All other stars inside the 3\,$\sigma$
  error circles lie further from the radio positions, and do not have
  blue $U_{300}-B_{390}$ colors. 

\begin{figure*}
\centering
\includegraphics[width=17cm, height=18cm]{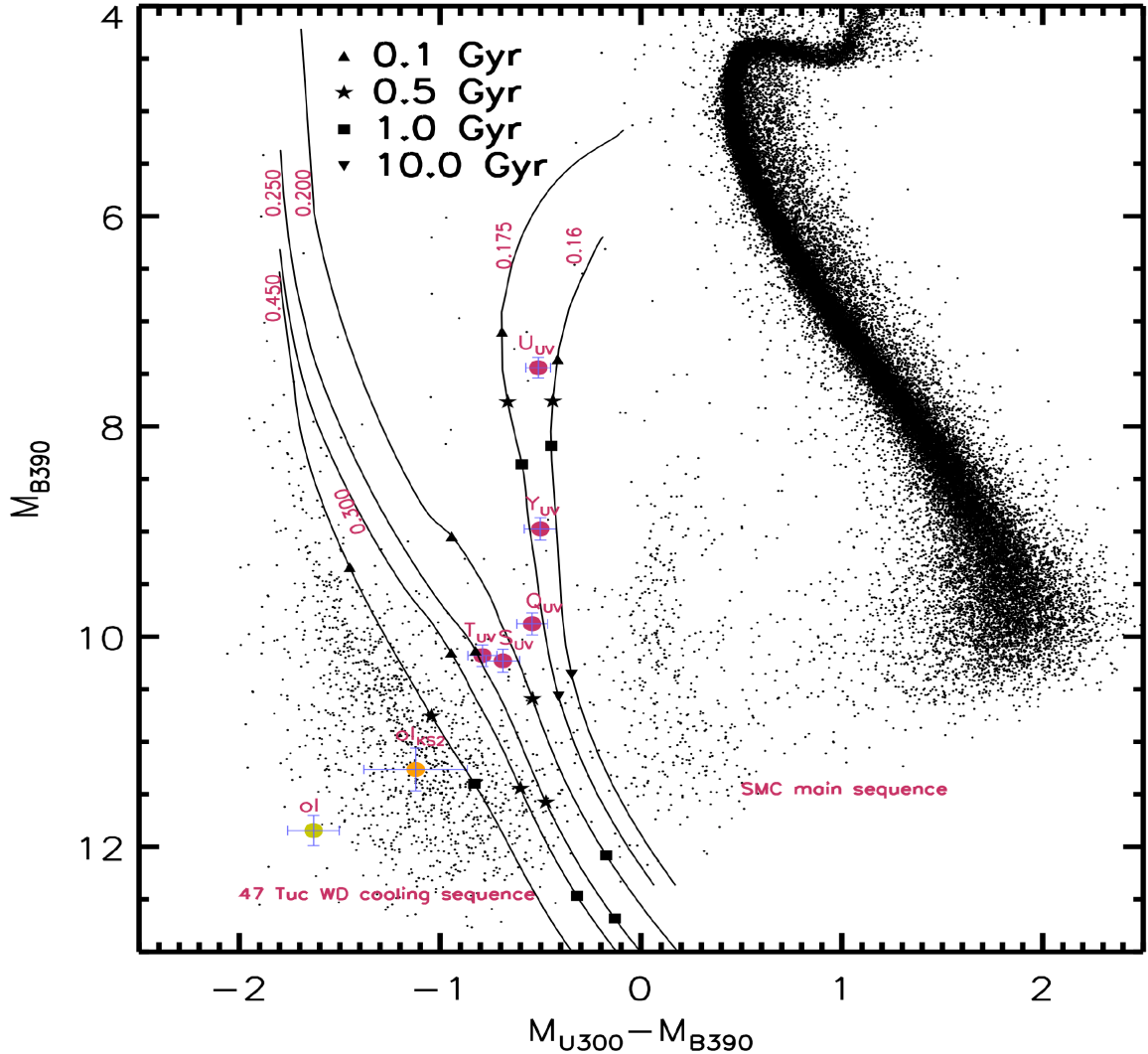}
\caption{$U_{300}-B_{390}$ versus $B_{390}$ CMD based on \change{DAOPHOT} photometry of our UVIS
  images. Each MSP companion is denoted with a letter and a filled magenta circle.
The green filled circle represents the blue object detected by DAOPHOT inside the 3\,$\sigma$ match circle of 47\,Tuc I. 
The orange filled circle is the same blue object detected by KS2 inside the 3\,$\sigma$ match circle of 47\,Tuc I. 
Calibrated magnitudes were converted to absolute magnitudes
  using a distance to 47\,Tuc of $4.69$ kpc \citep{2012-woodley} and
  reddening $E(B-V)=0.04$ \citep{2007-salaris}. The lines correspond to
  cooling models for He WDs with thick H envelopes as in  \citet{2002-serenelli}, computed for an initial metallicity of $Z=0.003$. Labels
  at the top of each track represent the mass of the He WD in units of solar mass.}
\label{fig:cmd}
\end{figure*}

\begin{figure*}
\centering
\includegraphics[width=18cm, height=12cm]{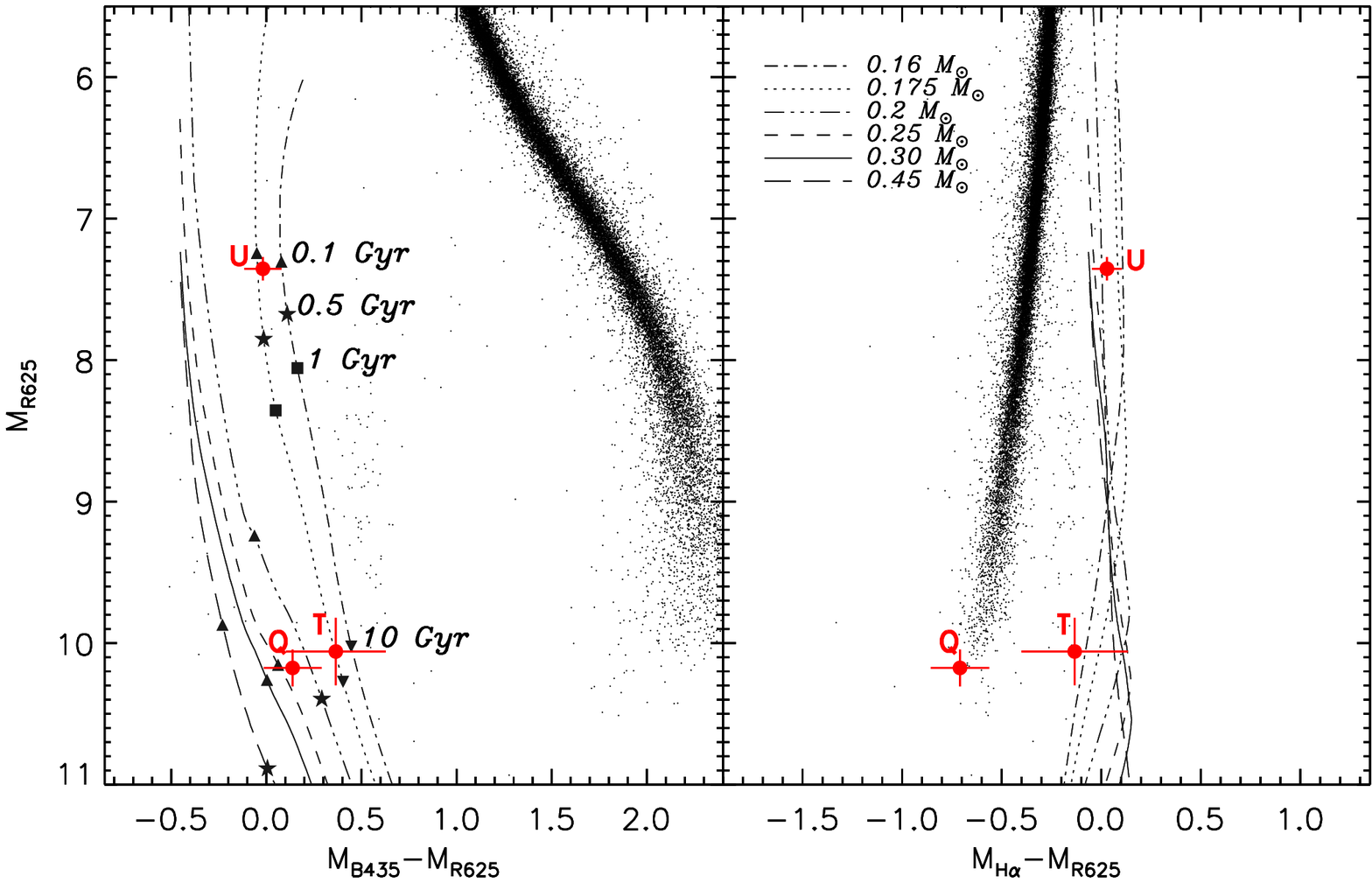}
\vspace{-20pt}
\caption{$B_{435}-R_{625}$ versus $R_{625}$ (left) and H$\alpha-R_{625}$
versus $R_{625}$ (right) CMDs from our WFC \change{DAOPHOT} photometry. MSP
counterparts are plotted with filled red circles. Calibrated Vega-mag
magnitudes were converted to dereddened absolute magnitudes 
\change{using a distance to 47\,Tuc of $4.69$ kpc \citep{2012-woodley} and reddening $E(B-V)=0.04$ \citep{2007-salaris}. 
We include He WD cooling tracks (Serenelli et al. 2002) for masses in the range 0.16 -- 0.45 $M_{\odot}$ and 
an initial metallicity of $Z=0.003$.} 
Photometry errors include the DAOPHOT
uncertainties, and errors in the distance modulus and $E(B-V)$.}
\label{fig:halpha} 
\end{figure*}

The blue stars that match with
  47\,Tuc T, Y and U are the likely WD counterparts proposed
  by E01, E03a and \citet{2006-bog}, while for 47\,Tuc Q and S these are new
  discoveries. Based on the similar locations in the $B_{390}$ versus
  $U_{300}-B_{390}$ color magnitude diagram (CMD), and the good
  alignment with the radio positions, we consider the stars near
  47\,Tuc Q and S to be the likely counterparts, and WD companions, to
  these MSPs, as well. 
   A more detailed discussion of the individual
  systems is given in Sections \ref{Q} to \ref{mspu}. In the remainder of the
  text we use capital letters (without subscript) to refer to the MSPs
  detected at radio wavelengths, while we add the subscript "UV" to
  refer to the NUV counterparts, e.g.\,47\,Tuc Q versus Q$_{\rm
  UV}$. Finding charts in the F300X and F390W filters are presented in
  Figure \ref{fig:findings}. Figures \ref{fig:cmd} and \ref{fig:halpha} are NUV and optical CMDs for 47\,Tuc that
  show the photometry for Q$_{\rm UV}$, S$_{\rm UV}$, T$_{\rm UV}$,
  U$_{\rm UV}$ and Y$_{\rm UV}$. In Table \ref{table:radio} we summarize the properties
  of the five MSPs as determined from radio and X-ray observations,
  while our astrometric and photometric results can be found in
  Table \ref{tab:offset}.

We have estimated the parameters of Q$_{\rm UV}$, S$_{\rm UV}$,
  T$_{\rm UV}$, U$_{\rm UV}$ and Y$_{\rm UV}$ assuming that they are
  He WDs. In Figures \ref{fig:cmd} and \ref{fig:halpha} we have included \change{new
evolutionary tracks for He WD cooling models as described in \citet{2002-serenelli},} but
with a metallicity as appropriate for 47\,Tuc of $Z=0.003$ \citep[\protect{$[Fe/H]=-0.78$,}][]{ 2014-Thygesen}. 
 Masses were estimated
  by determining the mass range that corresponds to the two cooling
  tracks that bracket the location of the companion in the NUV CMD,
  including $1\,\sigma$ error bars. 
Errors include the DAOPHOT uncertainties, errors in the
       calibration, and in the adopted distance modulus and extinction
       (see Section \ref{Photometry}).
The corresponding ranges in age, effective
  temperature T$_{eff}$, surface gravity $\log g$, and bolometric
  luminosity $L$ were determined by interpolating linearly along the
  two bracketing tracks using the absolute magnitude in the F390W
  filter ($M_{B_{390}}$) as the independent variable. Table \ref{table:ages}
  summarizes the results.

Cooling ages derived from \change{our} models (column 3 in Table \ref{table:ages})
  represent the time since the point of highest temperature in the
  cooling track until the observed location along the track. Prior to
  this time lapse is the time between Roche-lobe detachment (when the
  mass transfer from the He WD progenitor to the neutron star stops)
  until the point where the highest T$_{eff}$ is reached. This phase lasts
  only a few Myr for masses 0.25 $M_{\odot}$ and higher, but can
  amount to $\sim$1.5 Gyr for the 0.16 $M_{\odot}$ model, $\sim$1 Gyr
  for 0.175 $M_{\odot}$, and $\sim$500 Myr for 0.2 $M_{\odot}$. The exact value depends on
  the uncertain details of the binary evolution, and sets a lower limit to the total age.
 \citet{2014-Istrate} have derived
  a general expression for the duration of this "proto-WD" phase
  ($\Delta t_{\rm proto}$) in terms of the He WD mass, which for our MSP
  companions is given in the last column of Table \ref{table:ages}. 
We find that the proto-WD time scale in \change{our} models and the value of
  $\Delta t_{\rm proto}$ are in general agreement.

We have estimated the number of He WD-like objects that we
    expect to find by chance within a separation of 0\farcs016
    (i.e. the largest offset among the suggested matches for 47\,Tuc
    Q, S, T, U, and Y; see Table \ref{tab:offset}) from the radio positions. To this
    end, we determined the density of blue objects (stars
    arcsec$^{-2}$) enclosed by the cooling tracks for masses $0.16 \leq
    M_{WD}/M_{\odot} \leq 0.30$ and absolute magnitudes $6 \leq
    M_{B_{390}} \leq 11$ (see Figure \ref{fig:cmd}). For the total area of the
    3\,$\sigma$ match circles of nineteen MSPs, the expected number is
    very small, $9.5\times10^{-5}$. This suggests that our five
    identified MSP counterparts are indeed not chance coincidences.

\begin{figure}
\centering
\twofigbox{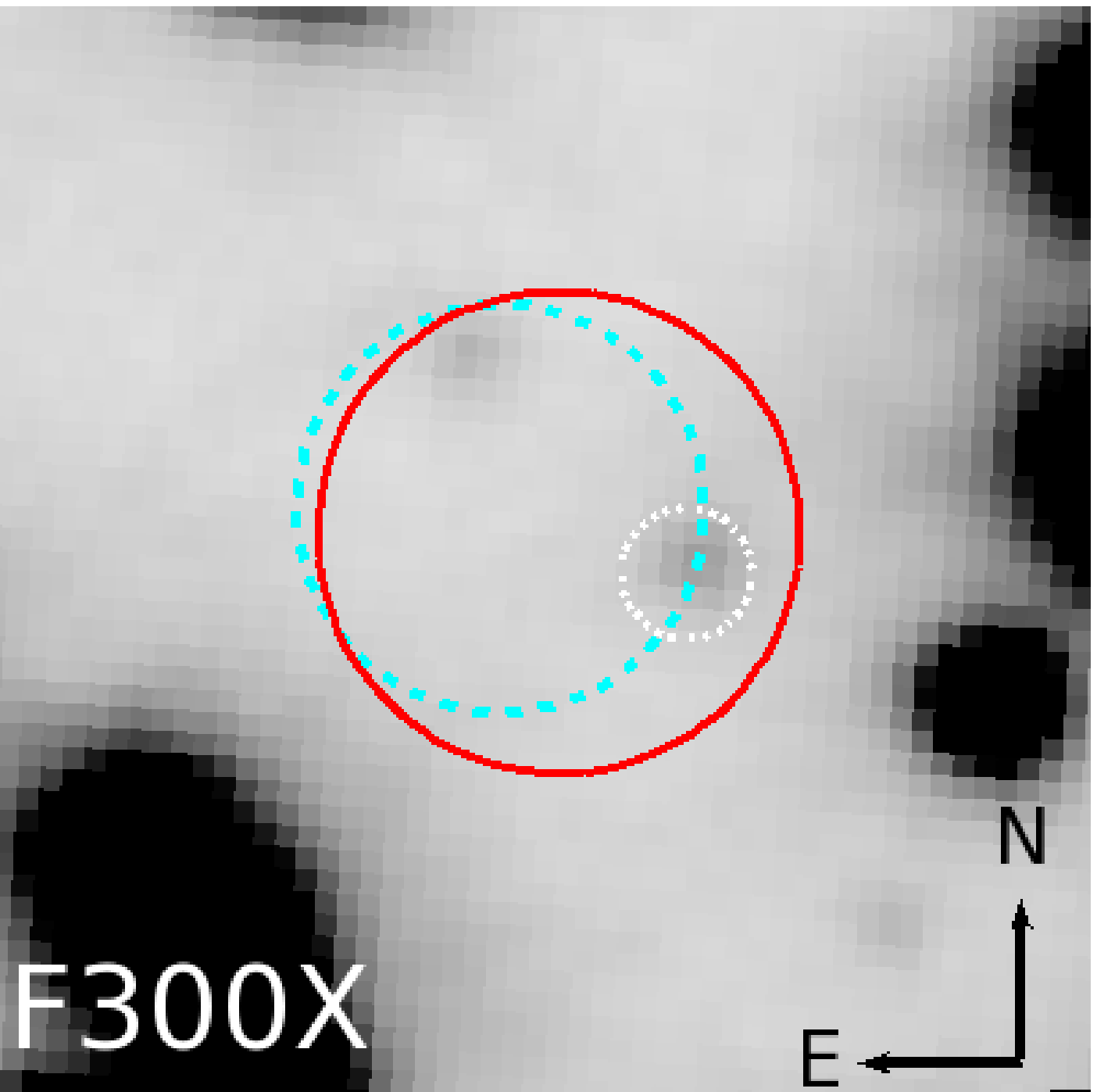}{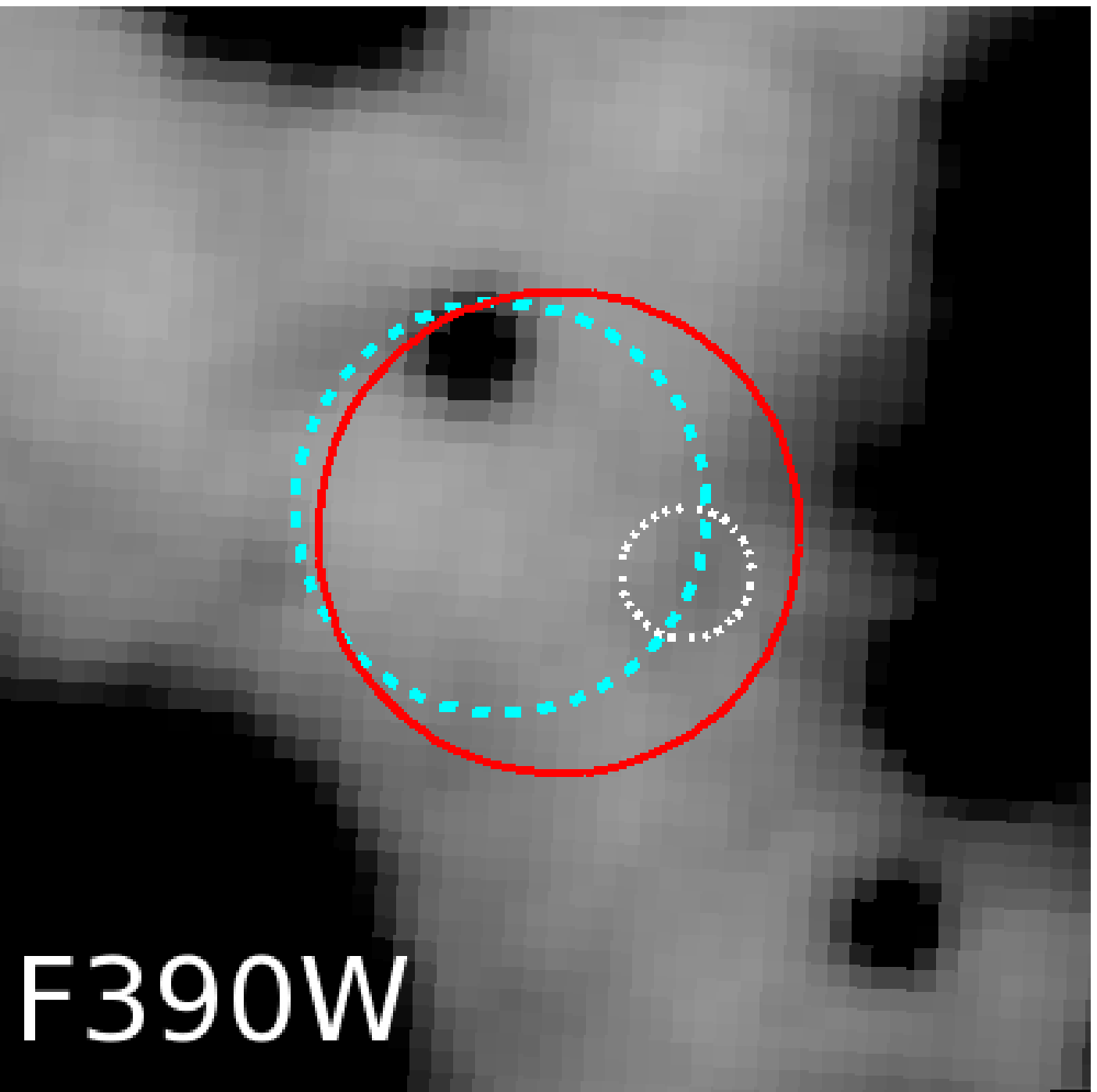}
\caption{Finding chart of 47\,Tuc I in the NUV. The solid red circles are the 3\,$\sigma$ match circles for the radio position of the pulsar.  They are centered on 47\,Tuc I. 
The dashed cyan circles are the 3\,$\sigma$ match circles of the {\em Chandra} source. The dashed white circles indicate the blue object that lies at 0\farcs12 from the pulsar. The images are $1\farcs0\times1\farcs0$ in size.}
\label{fig:mspI}
\end{figure}

For the binary MSP 47\,Tuc I we find a very blue object at 0\farcs12
  (1.6\,$\sigma$) with a $U_{300}-B_{390}$ color that puts it
  on or very close to the CO WD sequence (Figures \ref{fig:cmd} and \ref{fig:mspI}). This offset is
  much larger, and its color much bluer, than those for the five binary
  MSPs discussed above. We estimate the probability that this source
  is a spurious match by considering
        that blue ($U_{300}-B_{390} \lesssim -0.65$) objects are found in the
  error circles of two (47\,Tuc D and N) of the seven isolated MSPs in
  the UVIS field. Due to their intrinsic faintness in the NUV/optical,
  no counterparts are expected for the isolated MSPs. 
The blue objects close to 47\,Tuc D (at 0\farcs14 or
        1.8\,$\sigma$, and at (--1.44, 11.50) in the CMD of Figure \ref{fig:cmd}) and
        47\,Tuc N (at 0\farcs16 or 2.1\,$\sigma$, and at (--0.65, 9.68) in the
        CMD)---are therefore almost
  certainly chance alignments. The probability to find a blue star in
  the 3$\sigma$ error circle by chance in a single trial is therefore,
  approximately, $2/7\approx0.29$. This results in a (binomial)
  probability to find a blue random match for exactly one of ten\footnote{For
        47\,Tuc U and W excellent counterparts have already been
        identified.} binary MSPs in our field of $\sim$14\%. While we consider the NUV identification
  of 47\,Tuc I not very likely, we give more details on this object
  in Section 4.6, and provide a finding chart in Figure \ref{fig:mspI}.

\begin{table*}
\centering
  \caption{\change{Limits on companion masses 
($m_c$; for an assumed neutron star mass of 1.35 $M_{\odot}$) 
as derived from the mass function (Freire et al. in preparation).} Masses ($M_{WD}$), cooling ages, effective
temperatures (T$_{eff}$), surface gravities (log $g$) and luminosities ($L$) for the MSP companions
studied, as derived from our NUV photometry and \change{our} $Z=0.003$ He WD
cooling models. \change{In column 8, the inclination angle ($i$) as inferred from the mass function and the estimated masses $M_{WD}$.} 
The last column gives $\Delta t_{proto}$, 
which is the lapse of time from the Roche-lobe detachment until the proto-He WD reaches the WD cooling track, 
calculated using the relation derived by \citet{2014-Istrate}.}
  \begin{tabular}{c c c c c c c c c c}
  \hline
Companion & \change{$m_c$ ($M_{\odot}$)}&$M_{WD}$ ($M_\odot$)& Age$^1$ (Gyr)& T$_{eff}$ (kK) & log $g$ & log $L$ ($L_{\odot}$) & \change{$i$ ($^{\circ}$)}& $\Delta t_{proto}$ (\change{Gyr})\\
 \hline
 $\text{Q}_{\rm UV}$ &$>0.18$ & $0.175-0.20$   & $5.2-0.3$ &$8.9-10.6$ & $6.4-6.7$ &-1.9 $-$ -1.8& 90--63&$1-0.4$ \\
 $\text{S}_{\rm UV} $&$>0.09$ & $0.20-0.25$    & $0.4-0.1$ & $9.7-11.3$ & $6.7-7$  & -2.0 $-$  -1.9 & 28--22& $0.4-0.1$\\
$ \text{T}_{\rm UV}$ &$>0.17$ & $0.20-0.30$    & $0.4-0.1$ & $9.9-12.8$ & $6.7-7.2$ & -2.0 $-$ -1.8 &58--36&$0.4-0.02$\\
$\text{U}_{\rm UV}$  &$>0.12$ &  $0.16-0.175^2$& $0.2-0.3$ &$10.1-12$ & $5.5-5.8$ & -0.9 $-$ -0.8&51--46&$1.9-1.3$\\
$\text{Y}_{\rm UV}$  &$>0.14$ &  $0.16-0.175$  & $2.1-1.8$ &$8.9-10$     & $6-6.2$     & -1.6 $-$ -1.5&60--53&$1.9-1.3$\\
\hline
\label{table:ages}
\end{tabular}\\
\vspace{-12pt} 
{\flushleft
\startfoot  
\textfoot{Ages represent the time between the point of highest T$_{eff}$ on the
cooling track until the observed point along the track.}
\textfoot{An upper mass limit of $0.17$ $M_\odot$ was found by E01, for which they estimated an age of $\sim0.6$ Gyr, T$_{eff}=11\,000$ K and  log $g=5.6$ .}
}
\end{table*}

  For 47\,Tuc E, H, J (binaries), and L (isolated) not a single NUV
  object was detected inside their 3\,$\sigma$ match circles. For
  47\,Tuc O, R, and W (binaries) and C, F, G, and M (isolated) the
  objects in the match circles have colors that place them on the main
  sequence or sub-giant branch of 47\,Tuc and lie at least
  0\farcs11 (1.4\,$\sigma$) away from the radio positions. Without any
  additional information we cannot conclusively say if any of these
  are likely counterparts, and we do not discuss them further in this
  paper. We note that the true, non-degenerate, counterpart to 47\,Tuc
  W \citep{2002-Edmonds} is not detected in the NUV images. This is not unexpected as
  it is relatively faint (mean $V=22.3$) and red (compared to Q$_{\rm
  UV}$, S$_{\rm UV}$, T$_{\rm UV}$, U$_{\rm UV}$, and Y$_{\rm UV}$).

\subsection{47\,Tuc Q}
\label{Q}

In the CMD of Figure \ref{fig:cmd}, the counterpart of 47\,Tuc Q is located
  between the main sequence of the Small Magellanic Cloud (SMC) and
  the CO WD cooling sequence, at a similar $U_{300}-B_{390}$ color as
  U$_{\rm UV}$, but about about 2.4 mag fainter in $B_{\rm 390}$.
E03a reported the presence of a variable
 main-sequence turnoff star very close (at 0\farcs24)
 to the position of 47\,Tuc Q. This star, which they labelled "nQ",
 is the bright object to the south-west of Q$_{\rm UV}$ in Figure \ref{fig:findings}.
They discarded nQ as the true counterpart to the MSP, since its
 period of optical variability does not match the orbital period of the pulsar derived
 from radio timing. The faint blue star Q$_{\rm UV}$ lies
 closer to the radio position than nQ (0\farcs012 or $0.2\sigma$), but was not detected by E03a. It
 lies in the PSF wings of two stars (nQ and a star on the 47\,Tuc
 sub-giant branch), which makes Q$_{\rm UV}$ difficult to detect at
 optical wavelengths.

 In our $R_{625}$ versus $B_{435}-R_{625}$ CMD, Q$_{\rm UV}$ also
 lies in the region where He WDs are expected. But unlike the
 counterparts to 47\,Tuc U and T that lie to the red side \change{(H$\alpha$ faint side)} of the main
 sequence in the $R_{625}$ versus H$\alpha-R_{625}$ CMD (see
 Sect. \ref{mspT} and \ref{mspu}, and Figure \ref{fig:halpha}), there is no indication for the
 presence of a broad H$\alpha$ absorption line in Q$_{\rm UV}$: its
 H$\alpha -R_{625}$ color is consistent with the color of most other
 stars in the field with similar $R_{625}$ magnitude. If Q$_{\rm UV}$ actually
 is a He WD, the H$\alpha-R_{625}$ color could be explained if
 there is some residual H$\alpha$ emission in the system that
 effectively fills in the H$\alpha$ absorption line of the H-rich
 outer layer of the WD. From the location of Q$_{\rm UV}$ in
 a $B_{625}-R_{625}$ versus H$\alpha-R_{625}$ color-color diagram
 and synthetic colors of simulated spectra, we estimate that the
 equivalent width of such an emission line is about $-70$ \AA\,.
However, the inclination inferred from the mass function (Table \ref{table:radio}) and our 
estimate for $M_{WD}$ (Table \ref{table:ages}) is high (\change{63}$^\circ$ to 90$^\circ$), so eclipses are
expected if a large cloud of ionized material would be present. 
The absence of radio eclipses in 47\,Tuc Q then argues against the presence 
of such an extended H$\alpha$ emission region.
 We note that, given the proximity of the
 two brighter stars, the photometry for Q$_{\rm UV}$ should be
 considered with some caution; in particular, the PSFs of relatively
 bright stars in the F625W and H$\alpha$ images shows Airy disk
 peaks around the central region
 of the PSF, which coincide with the location of Q$_{\rm UV}$
 and could affect its colors if not properly modelled.

The location of Q$_{\rm UV}$ with respect to the He WD cooling
 tracks suggests a mass between $\sim$0.175 and 0.20
 $M_{\odot}$. This is consistent with the lower limit of the
 companion mass derived from the pulsar mass function (0.18
 $M_{\odot}$ for an assumed neutron star mass of 1.35 $M_{\odot}$). The
 derived cooling age is very uncertain, as Q$_{\rm UV}$ lies in between the
 tracks for the slowly cooling 0.175 $M_{\odot}$ models and the
 rapidly cooling 0.20 $M_{\odot}$ He WDs; these respective masses
 predict an age range of $5.2-0.3$ Gyr. In addition, the
 proto--WD phase can add up to 1 Gyr to obtain the time that passed
 since Roche lobe detachment. However, a comparison of the WD age with the characteristic pulsar spin-down age $\tau_c$ 
 is difficult. The observed period derivatives for the 47\,Tuc MSPs
 are dominated by the effect of the acceleration of the pulsars along
 the line of sight in the gravitational potential of the cluster. As
 a result, the intrinsic period derivatives,
and therefore the pulsar
 characteristic ages, are poorly constrained. However, better estimations
of the intrinsic period derivatives, for stable MSP-WD systems, can be made by using the time derivative of the observed orbital periods, which
essentially measures the acceleration of each binary MSP along the line of sight. Details about this technique will be given in Freire et al. (in preparation). 
By using this method we obtain $\tau_c > 1.43$ Gyr for Q$_{\rm UV}$.

\subsection{47\,Tuc S}

S$_{\rm UV}$, together with T$_{\rm UV}$, is the faintest MSP
  companion that we found. This object is clearly visible in our F435W
  images, but in the F625W and H$\alpha$ images a reliable detection
  cannot be made. This is partly the result of a relatively bright
  diffraction spike of a brighter star that passes within a few pixels
  of the center of the PSF of S$_{\rm UV}$. From comparison with the
  He WD cooling tracks, we derive a mass of about $0.2-0.25$
  $M_{\odot}$ and an age of about $0.4-0.1$ Gyr. Taking into account its time spent as a "proto-WD"
  (up to 0.4 Gyr), this age range is roughly consistent with the lower
  limit on the pulsar characteristic age.

47\,Tuc S and the isolated pulsar 47\,Tuc F are only $0\farcs74$ apart. Both are associated with a single
  {\em Chandra} source, W\,77, which is likely a blend of (at least)
 these two sources. This explains the relatively large offset between
  S$_{\rm UV}$ and W\,77 in Table \ref{tab:offset}.

\subsection{47\,Tuc T}
\label{mspT}

We derive a mass for T$_{\rm
    UV}$ of $0.2-0.3$ $M_{\odot}$, in agreement with the lower limit of
    the companion mass (Table \ref{table:radio}). The approximate total WD age is about $0.8-0.1$ Gyr.
This value is roughly consistent with the lower limit on $\tau_c$ of this MSP.

T$_{\rm UV}$ lies in between the CO WD cooling sequence and the SMC
  main sequence in our NUV CMD.  The blue color of T$_{UV}$ is apparent from the finding charts in Figure \ref{fig:findings}.
  In the F390W image T$_{\rm UV}$ has a close neighbor at a separation
  of only 0\farcs07. In the F300X image the
  neighbor, as well as stars of similar F390W magnitude, are much
  fainter. T$_{\rm UV}$ is clearly
  detected in our F435W images. While also detected in F625W
  and F658N, it is very faint at these redder wavelengths; the
  close neighbor and the contamination of its PSF by a weak
  diffraction spike of another neighboring star, leads to relatively
  large errors in its optical colors (Figure \ref{fig:halpha}). Nevertheless, the
  optical counterpart is also clearly blue, and its location on the
  red side of the H$\alpha-R_{625}$ main sequence points at the
  presence of a strong H$\alpha$ absorption line. T$_{\rm UV}$ was
  already identified as the likely counterpart to 47\,Tuc T by E03a.
They detected T$_{\rm UV}$ in their $U$ images but
  not in $V$ and $I$, and could therefore only derive an upper limit
  to the color.

\subsection{47\,Tuc Y}

Our photometry confirms the blue color of Y$_{\rm UV}$ reported
    by E03a. We estimate that the mass of Y$_{\rm UV}$ is between
    0.16--0.175 $M_{\odot}$, in agreement with the lower limit from
    the radio mass function (Table \ref{table:radio}). The total age estimate, about 2
    Gyr for the cooling time plus $1.3-1.9$ Gyr for the proto-WD phase,
    makes Y$_{\rm UV}$, possibly together with Q$_{\rm UV}$, the oldest of
  the five WD companions discussed in this paper. 

Y$_{\rm UV}$
 is clearly detected in our F435W image. The PSF of the
 neighboring bright object, which can be seen a little offset from the
 center in the finding charts of Figure \ref{fig:findings}, reduces the sensitivity
 for very faint objects. As a result, Y$_{\rm UV}$ is not detected in
 the $R_{625}$ and H$\alpha$ images.

\subsection{47\,Tuc U}
\label{mspu}

E01 found that U$_{\rm UV}$ is a
    variable blue object, with a semiamplitude variation of 0.004 mag and with a period that is consistent with the orbital period derived from radio timing (about 0.43 days). 
Based on a comparison between the $U$, $V$, and $I$
    photometry for this star and the \citet{2001-serenelli} models
    for He WD evolution, its mass and age were estimated to be
    $\sim$0.17 $M_{\odot}$ and $\sim$0.6 Gyr, respectively.
E01 also identified this object as the counterpart of the X-ray source W11.
 E01 indeed found that the He WD companion is well inside
    its Roche lobe, which also indicates there is no ongoing accretion
    that contributes to the X-rays.

U$_{\rm UV}$ is the brightest of the five
    counterparts discussed in this work. It is an isolated object, so its
    UVIS and WFC photometry is robust. U$_{\rm UV}$ is located above and to the left of the SMC main sequence 
in the NUV and optical CMDs, in agreement with the results of E01 and E03a. 
The location of U$_{\rm UV}$ to the red side of the main
    sequence in the $R_{625}$ versus H$\alpha-R_{625}$ CMD points at
    the presence of a strong H$\alpha$ absorption line, and is indeed
    in agreement with the He WD cooling tracks.

The He WD atmosphere models \change{used in this paper are computed
    for a lower metallicity and make use of updated atmosphere models \citep{2011-rohr}
with respect to the ones used in E01}. 
 From the updated models and our UVIS photometry, we
    constrain the mass for U$_{\rm UV}$ to be $0.16-0.175$ $M_{\odot}$,
    T$_{eff} \approx 11,000$ K, and $\log g \approx 5.6$. This is
    consistent with the results from our WFC photometry and with the
    parameters derived by E01. Our cooling age estimate ($0.2-0.3$
    Gyr) is somewhat lower than found in the latter study ($\sim$0.6
    Gyr). If we take into account $\Delta t_{proto}$, the total age
     becomes 2.1--1.6 Gyr; this is still consistent with the
     lower limit to $\tau_c$ from Table \ref{table:radio}.

\subsection{47\,Tuc I}

  The blue star in the 3\,$\sigma$ error circle of 47\,Tuc I is very
  faint. The errors in the photometry are considerable, and we plot
  the results from both DAOPHOT and KS2 in Figure \ref{fig:cmd} (denoted as oI and oI$_{\rm KS2}$, respectively); it is too faint
  to be detected in any of our optical images. This object is too blue
  to be a He WD. While the colors are more consistent with a CO WD
  than a He WD, the former option is not likely if the star is truly
  associated with 47\,Tuc I. MSPs with CO WD companions typically have
  longer spin periods compared to systems with He WD companions
(\citeauthor {2012-Tauris-langer} \citeyear{2012-Tauris-langer}; 
note the exception of PSR\,J1614--2230). The
  spin period of 47\,Tuc I, on the other hand, is very short
  ($P_s=3.2$ ms; Freire et al. 2003). We also point out that the mass function suggests
  that the companion is a very low-mass object, with $m_c > 0.013$
  $M_{\odot}$. If the blue object is the $\sim0.5$ $M_{\odot}$ CO WD
  companion to 47\,Tuc I, this would imply an inclination of
  $i\approx1.8^{\circ}$. None of the aforementioned arguments are
  conclusive grounds for rejecting or confirming the association.

\section{Discussion}

The five binary MSPs in 47\,Tuc that we discuss in this work are
characterized by spin periods below 8 ms, orbital periods $P_b$
between 0.43 and 1.2 d, and companions more massive than
$\sim$0.09 $M_{\odot}$ (Table \ref{table:radio}). Based on a compilation of available
data for binary pulsars in the Galactic disk, the nature of the
companion for such binary pulsars is most likely a He WD \citep[see][]{2012-Tauris-langer}. 
We find that our NUV/optical photometry for the five MSP
counterparts is indeed consistent with the magnitudes and colors
predicted by the low-metallicity He WD cooling models by \citet{2002-serenelli}. 
In particular, the derived mass constraints lie in the
range $0.16-0.30$ $M_{\odot}$. Our results for the companion of 47\,Tuc
U are consistent with those derived by E01, who
compared $U$, $V$, and $I$ photometry for this star with the
solar-metallicity He WD models by \citet{2001-serenelli}.

The standard formation scenario for an MSP accompanied by a He WD
involves a long period of stable mass transfer. In the models by
\citet{2002-serenelli}, the He WDs emerge from this phase with a
thick H envelope.  Calculations for the subsequent evolution of white
dwarfs predict a dichotomy in cooling behavior\footnote{The evolutionary models by \citet{2014-Istrate} do not
  show this dichotomy in cooling behavior depending on the occurrence
  of H shell flashes or not, but rather indicate that the cooling
  timescale mainly depends on the mass of the proto--WD.}
\citep[e.g.][]{1996-Alberts,2001-althaus,2001-serenelli,2007-Panei,2013-Altahus}. 
He WDs with masses $\gtrsim$0.2 $M_{\odot}$ 
\citep[\protect{or equivalently
$P_b\gtrsim1.55$ d; see}][]{2005-vank} experience
thermonuclear H-shell flashes before they reach the cooling
track. 
These flashes leave the WD with a thin H envelope, as a result of additional
diffusion-induced H-shell flashes. After these short-lived episodes,
the WD remnant cools down on a relatively short timescale. In He WDs less
massive than $\sim$0.2 $M_{\odot}$ H-shell flashes do not occur, and
therefore a thick H envelope is retained. Residual H burning can keep
up a high temperature over long timescales, resulting in much longer
cooling times. Figure \ref{fig:cmd} implies that S$_{\rm UV}$ and T$_{\rm UV}$
cool fast, U$_{\rm UV}$ and Y$_{\rm UV}$ cool slowly, while for
Q$_{\rm UV}$ the time scale is ambiguous. Unfortunately, the spin-down
ages of the pulsars in 47\,Tuc are too uncertain to give an
independent constraint on the WD age, and hence, mass.

There are only two binary MSPs in our field whose mass functions
    suggest He WD-like companions but for which we have not found a
    plausible counterpart, viz.~47\,Tuc E and H. Their orbital periods
    are the longest of the binary MSPs in 47\,Tuc included in our
    field ($\sim$2.3 d). The relation between orbital period and
    companion mass \citep[see for example][]{1999-tauris} then suggests that the companions in these systems
    are more massive than those in the other MSPs. Since a higher mass
    implies a shorter cooling time, the non-detection of the
    companions to 47\,Tuc E and H can be understood if they were
    formed more than $\sim$0.7 Gyr ago so that they have dimmed below
    the detection limit.

\begin{figure}
\centering
\includegraphics[width=0.47\textwidth]{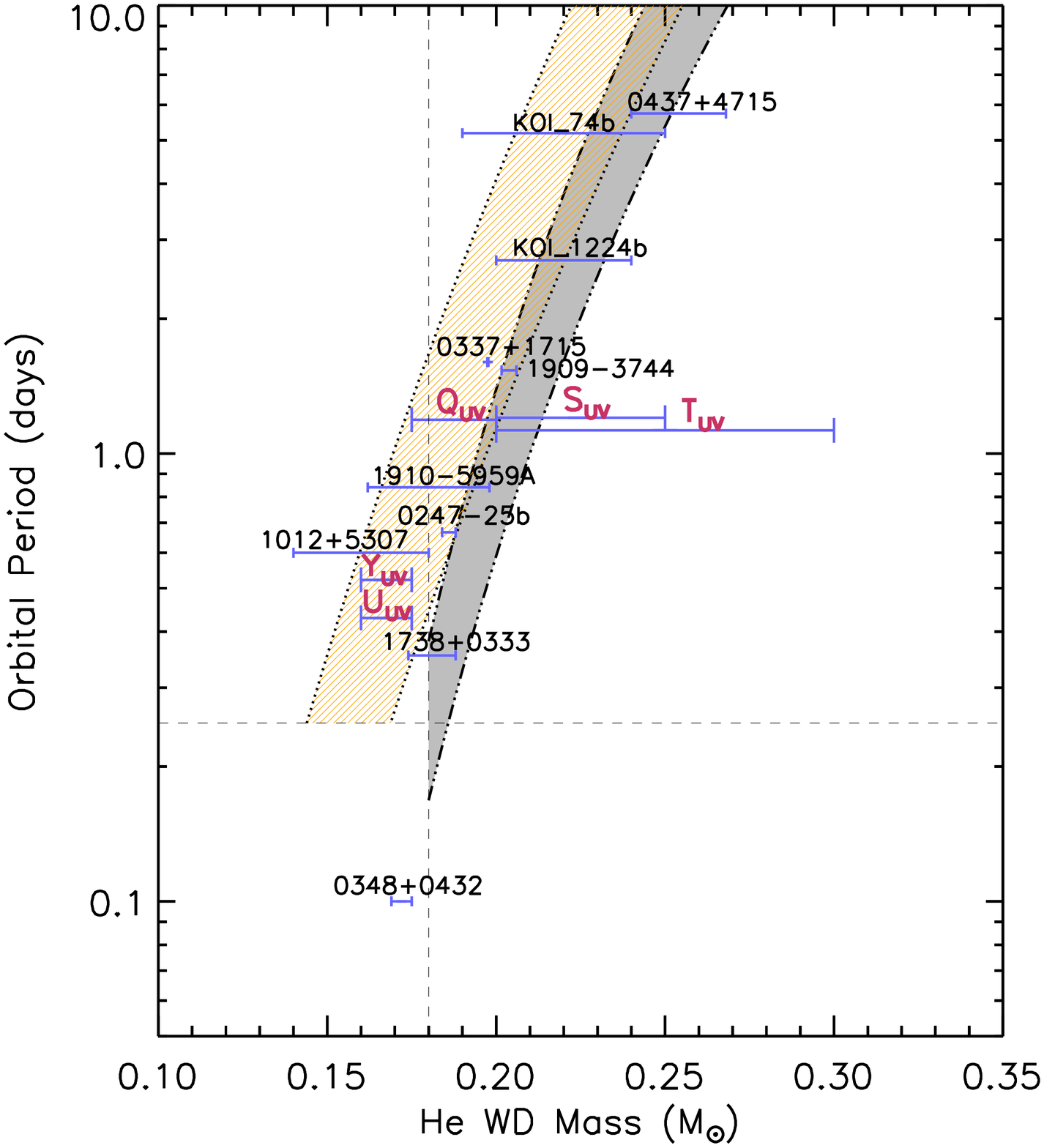}
\vspace{-5pt} 
\caption{Mass-orbital period relations for He WDs. In gray is the relation as calculated by \citet {1999-tauris},
with metallicities in the range $Z=0.001-0.02$ (from right to left) and for WD masses in the range $0.18-0.45$ $M_{\odot}$. In orange the relation by
\citet{2010-vito} for solar metallicity and for orbital periods $>0.25$ d. The width of the relation corresponds to the uncertainties in its parameters.
Overplotted are our results for the companions of 47\,Tuc Q, S, T, U and Y. 
The error bars denote the possible masses for each object as determined from the WD cooling tracks by \citet{2002-serenelli}.
We also plot other pulsar companions for which $1\sigma$ uncertainties are less than $10\%$ of the WD mass (see \citealt{2014-taurisvan} and references therein).
The dashed vertical line denotes the lower mass limit (0.18 $M_{\odot}$) for which the \citet {1999-tauris} relation is valid.
The dashed horizontal line shows the minimum orbital period (0.25 d) for which the \citet {2010-vito} relation is valid.
}
\label{fig:massperiod}
\end{figure}

In Figure \ref{fig:massperiod} we show the relation between the mass of the He WD and the
orbital period based on binary-evolution calculations by \citeauthor{1999-tauris} (\citeyear{1999-tauris}; valid for He WDs masses between 0.18 and 0.45
$M_{\odot}$) and  \citet{2010-vito}, valid for $P_b>0.25$ d. 
We find that the orbital periods and mass estimates for
S$_{\rm UV}$ and T$_{\rm UV}$ are more consistent with the relation derived by \citet{1999-tauris}. In contrast,
Y$_{\rm UV}$ and U$_{\rm UV}$ are offset from that
relation, but their estimated masses lie below the mass range to which
the relation applies. However, we observe that Y$_{\rm UV}$ and U$_{\rm UV}$
are in good agreement with the relation by \citet{2010-vito}. Q$_{\rm UV}$ is consistent with both relations.
 \citet{2010-vito} made different assumptions on the mass transfer process and they obtain, for the same orbital periods, lower masses for the WD companions than \citet{1999-tauris}. They
also find that their mass -- orbital period relation is not sensitive to the initial mass of the accreting neutron star. 
Studies about the assumption on how conservative the mass transfer is, 
have found that the dispersion  around the 
\citet{2010-vito}  relation decreases as the He WD mass increases \citep{2012-vito}. 

The five binary MSPs discussed here have X-ray luminosities $L_X= 2.9-8.9\times 10^{30}$ erg s$^{-1}$ \citep{2006-bog}. Like for
  the majority of the 47\,Tuc MSPs, it was found that their X-ray
  emission can be described by a thermal (black-body or neutron-star
  hydrogen atmosphere) spectrum as expected for emission from the
  heated magnetic polar caps of the neutron stars. Using the $L$, T$_{eff}$
  and mass estimates from Table \ref{table:ages}, and the orbital periods from
  Table \ref{table:radio}, we have calculated the WD radii (R$_{WD}$) and
  (assuming a neutron-star mass of 1.35 $M_{\odot}$) the corresponding
  radii of the Roche Lobe ($R_{L,WD}$; using the formula of \citealt{1971-pac}) . We indeed find that all five
  WD companions fit well within their Roche lobe ($R_{L,WD}/R_{WD}
  \gtrsim 5$) as already found for 47\,Tuc U by E01, 
 so we indeed expect that accretion does not contribute to the X-ray luminosity.

For the five systems we studied, we calculate the contribution of the MSP radiation to
the observed NUV luminosity of the WD. We used the values of the spin-down luminosity ($\dot E$) for each MSP
given by  \citet{2006-bog}.
Assuming isotropic emission from the neutron star we calculate the energy flux intercepted by the WD ($\dot E_{\rm i}$)
using our estimates of R$_{WD}$ and the separation between the stars.
For Q$_{\rm UV}$, S$_{\rm UV}$ and T$_{\rm UV}$ we obtain log\,$\dot E_{i}$ ($L_{\odot}$) $<-4$ which is well
below their NUV luminosities. This result indicates that irradiation 
is not driving the luminosities of these systems. 
On the other hand, for Y$_{\rm UV}$ and U$_{\rm UV}$
we obtain $-4.2 < $log\,$\dot E_{i}$ ($L_{\odot}$)$< -2.3$ which is a closer value to their respective $L$, implying that 
heating might be present, though not dominating the NUV
luminosity. This result can be understood if we take into account that these two companions
have larger R$_{WD}$ and are closer to their MSPs than the other three WDs. 

We note that all the companions have cooling ages $\lesssim6$ Gyr (including age uncertainties), 
which are considerably less than the age of the cluster \citep[\protect{$9.9\pm0.7$ Gyr,}][]{2013-hansen}. 
\citet{2000-rasio} proposed that the formation of recycled MSP systems happened mostly during the early life 
of the cluster. However, our results may suggest that dynamical interactions during the later evolution of the cluster
have a more important role in the formation of these systems \citep[for an evolutionary scenario in this context see for example][]{2008-fre}.

The orbital periods of the five binary MSPs studied in this paper
    are all longer than the time span of the NUV observations ($\sim$9
    h); light curves constructed from these data alone give incomplete
    phase coverage. We defer a variability study, based on these and
    other {\em HST} data sets, to a future paper.

\section*{Acknowledgments}

We appreciate the comments by the anonymous referee, which improved this work.
L.E Rivera-Sandoval acknowledges support from a CONACyT fellowship.

\bibliographystyle{mn2e}
\bibliography{ms-bibfile}

\appendix

\label{lastpage}

\end{document}